\documentclass[useAMS,usenatbib]{mn2e}
\usepackage{txfonts}
\usepackage{graphicx}
\usepackage{natbib}

\def\del#1{{}}

\newcommand{\ltsima}{$\; \buildrel < \over \sim \;$}
\newcommand{\lsim}{\lower.5ex\hbox{\ltsima}}
\newcommand{\gtsima}{$\; \buildrel > \over \sim \;$}
\newcommand{\gsim}{\lower.5ex\hbox{\gtsima}}
\newcommand{\bra}{\langle}
\newcommand{\ket}{\rangle}

\newcommand{\solar}{\sun}
\newcommand{\planck}{{\em Planck}}
\newcommand{\plancks}{{\em Planck}'s }

\newcommand{\mK}{{\em m}K}

\newcommand{\nK}{{\em n}K}
\newcommand{\hfi}{{\em HFI}}
\newcommand{\lfi}{{\em LFI}}
\newcommand{\dd}{\mathrm{d}}
\newcommand{\e}{\mathrm{e}}

\title[Detecting SZ-Clusters with \planck]
{Detecting Sunyaev-Zel'dovich clusters with \planck:\\ III. Properties of the expected SZ-cluster sample}
\author[Bj{\"o}rn Malte Sch\"afer and Matthias Bartelmann]
{Bj{\"o}rn Malte Sch\"afer$^{1,2}$\thanks{e-mail: Bjoern.Schaefer@port.ac.uk (BMS);
mbartelmann@ita.uni-heidelberg.de (MB)}
and Matthias Bartelmann$^{3}$\footnotemark[1]\\
$^1$Max-Planck-Institut f\"ur Astrophysik, Karl-Schwarzschild-Stra{\ss}e 1, Postfach 1317, 85741 Garching, Germany\\
$^2$Institute of Cosmology and Gravitation, University of Portsmouth, Mercantile House, Hampshire Terrace, Portsmouth PO12EG, United Kingdom\\
$^3$Institut f{\"u}r theoretische Astrophysik, Zentrum f{\"u}r Astronomie, Universit\"at Heidelberg, Albert-Ueberle-Stra{\ss}e 2, 69120 Heidelberg, Germany}

\begin{document}
\pagerange{\pageref{firstpage}--\pageref{lastpage}}
\pubyear{2003}
\maketitle
\label{firstpage}

\begin{abstract}
The \planck~mission is the most sensitive all-sky submillimetric mission currently being planned and prepared. Special emphasis is given to the observation of clusters of galaxies by their thermal Sunyaev-Zel'dovich (SZ) effect. In this work, the results of a simulation are presented that combines all-sky maps of the thermal and kinetic SZ-effect with cosmic microwave background (CMB) fluctuations, Galactic foregrounds (synchrotron emission, thermal emission from dust, free-free emission and rotational transitions of carbon monoxide molecules) and sub-millimetric emission from planets and asteroids of the Solar System. Observational issues, such as \plancks beam shapes, frequency response and spatially non-uniform instrumental noise have been incorporated. Matched and scale-adaptive multi-frequency filtering schemes have been extended to spherical coordinates and are now applied to the data sets in order to isolate and amplify the weak thermal SZ-signal. The properties of the resulting SZ-cluster sample are characterised in detail: Apart from the number of clusters as a function of cluster parameters such as redshift $z$ and total mass $M$, the distribution $n(\sigma)\dd\sigma$ of the detection significance $\sigma$, the number of detectable clusters in relation to the model cluster parameters entering the filter construction, the position accuracy of an SZ-detection and the cluster number density as a function of ecliptic latitude $\beta$ is examined.
\end{abstract}


\section{Introduction}
The Sunyaev-Zel'dovich (SZ) effects \citep{1972SZorig, 1980ARA&A..18..537S, 1995ARA&A..33..541R, 1993birkinshaw} are promising tools for detecting clusters of galaxies out to high redshifts by their spectral imprint on the cosmic microwave background (CMB). This paper compiles the results of an extensive assessment of the observability of the SZ-effect for the European \planck-surveyor satellite based on numerical data, as described two preceeding papers (Sch{\"a}fer et al. 2004a,b). In the simulation, we try to model as many aspects of a survey of the CMB sky with \planck~as possibly relevant to the search for Sunyaev-Zel'dovich clusters of galaxies and the extraction of the weak SZ-signal. We discuss shortcomings of the simulation and of the filtering schemes, quantify the properties of the resulting SZ-cluster sample and compare our results with previous studies.

This paper contains a recapitulation of basic SZ-quantities (Sect.~\ref{sect_szdef}) and a brief outline of the simulation (Sect.~\ref{sect_sim}). The key results are compiled (Sect.~\ref{sect_result}) with a subsequent discussion (Sect.~\ref{sect_summary}). The cosmological model assumed throughout is the standard \mbox{$\Lambda$CDM} cosmology, with the parameter choices: $\Omega_\mathrm{M} = 0.3$, $\Omega_\Lambda =0.7$, $H_0 = 100\,h\,\mbox{km~}\mbox{s}^{-1}\mbox{ Mpc}^{-1}$ with $h = 0.7$, $\Omega_\mathrm{B} = 0.04$, $n_\mathrm{s} =1$ and $\sigma_8=0.9$.

\section{The Sunyaev-Zel'dovich effects}\label{sect_szdef}
The Sunyaev-Zel'dovich effects are very sensitive tools to observe clusters of galaxies out to large redshifts in submillimetric data. Inverse Compton scattering of CMB photons off electrons of the ionised ICM produces a modulation of the CMB spectrum and gives rise to surface brightness fluctuations of the CMB. One distinguishes the thermal SZ-effect, where the thermal energy content of the ICM is tapped from the kinetic SZ-effect, where the CMB photons are coupled to the bulk motion of the ICM electrons, as opposed to the thermal motion of the electrons in the thermal SZ-effect.

The relative change $\Delta T/T$ in thermodynamic CMB temperature at position $\bmath{\theta}$ as a function of
dimensionless frequency $x=h\nu /(k_B T_\mathrm{CMB})$ due to the thermal SZ-effect is given by:
\begin{eqnarray}
\frac{\Delta T}{T}(\bmath{\theta}) & = & y(\bmath{\theta})\,\left(x\frac{e^x+1}{e^x-1}-4\right)\mbox{ with }\\
y(\bmath{\theta}) & = & \frac{\sigma_\mathrm{T} k_B}{m_\e c^2}\int\dd 
l\:n_\e(\bmath{\theta},l)T_\e(\bmath{\theta},l)\mbox{,}
\label{sz_temp_decr}
\end{eqnarray}
where the amplitude $y$ of the thermal SZ-effect is the thermal Comptonisation parameter $y$. $y$
is defined as the line-of-sight integral of electron density times electron temperature. $m_\e$, $c$,
$k_B$ and $\sigma_\mathrm{T}$ denote electron mass, speed of light, Boltzmann's constant and the Thompson cross section,
respectively. The kinetic SZ-effect arises due to the motion of the cluster parallel to the line-of-sight relative to the cosmological frame of reference given by the CMB:
\begin{equation}
\frac{\Delta T}{T}(\bmath{\theta}) = -w(\bmath{\theta})\;\mbox{ with }\;
w(\bmath{\theta}) = \frac{\sigma_\mathrm{T}}{c}\int\dd l\:n_\e(\bmath{\theta,l})\upsilon_r(\bmath{\theta},l)\mbox{.}
\end{equation}
$\upsilon_r$ is the radial component of the cluster's velocity. The amplitude $w$ of the kinetic SZ-effect is called
the kinetic Comptonisation. 

The SZ-observables are the line-of-sight Comptonisations integrated over the cluster face. The quantities $\mathcal{Y}$ and $\mathcal{W}$ are consequently called the integrated thermal and kinetic Comptonisations, respectively:
\begin{eqnarray}
\mathcal{Y} & = & \int\dd\Omega\: y(\bmath{\theta}) = 
d_A^{-2}(z)\,\frac{\sigma_\mathrm{T} k_B}{m_e c^2}\:\int\dd V\:n_e T_e,\\
\mathcal{W} & = & \int\dd\Omega\: w(\bmath{\theta}) = 
d_A^{-2}(z)\,\frac{\sigma_\mathrm{T}}{c}\:\int\dd V\:n_e \upsilon_r.
\end{eqnarray}
$d_A(z)$ denotes the angular diameter distance to the cluster at redshift $z$. The fluxes of the thermal SZ-effect 
$S_\mathcal{Y}(x)$ and of the kinetic SZ-effect $S_\mathcal{W}(x)$ as functions of observing frequency are given by eqns.~(\ref{eq:S_thSZ}) and (\ref{eq:S_kinSZ}), respectively. The flux density of the CMB has a value of $S_0=22.9~\mathrm{Jy}/\mathrm{arcmin}^2$:
\begin{eqnarray}
S_\mathcal{Y}(x) & = & S_0\,\mathcal{Y}\,
\frac{x^4\,\exp(x)}{(\exp(x)-1)^2}\,\left[x\frac{\exp(x)+1}{\exp(x)-1} - 4\right]\mbox{.}
\label{eq:S_thSZ} \\
S_\mathcal{W}(x) & = & S_0\,\mathcal{W}\,\frac{x^4\,\exp(x)}{(\exp(x)-1)^2}\mbox{.}
\label{eq:S_kinSZ}
\end{eqnarray}

Exemplarily, table~\ref{table_planck_channel} summarises the fluxes $S_\mathcal{Y}$ and $S_\mathcal{W}$ and the corresponding changes in antenna temperature $T_\mathcal{Y}$ and $T_\mathcal{W}$ for Comptonisations of $\mathcal{Y} = \mathcal{W} = 1~\mathrm{arcmin}^2$.

\begin{figure}
\resizebox{\hsize}{!}{\includegraphics{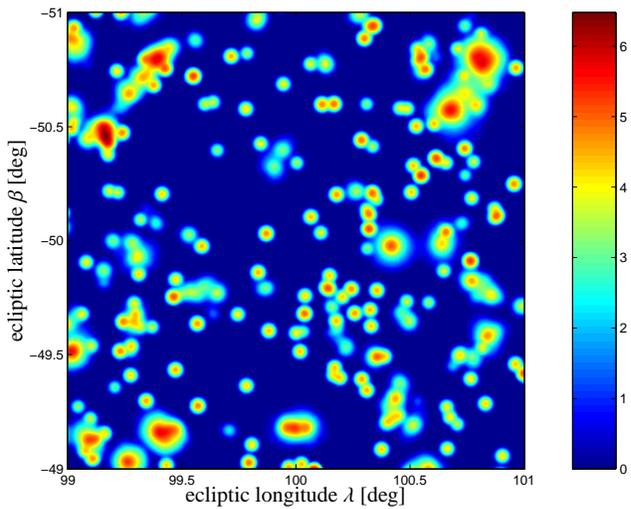}}
\caption{Detail of the thermal Comptonisation map: A $2\degr\times2\degr$ wide cut-out centered on the ecliptic 
coordinates $(\lambda,\beta) = (100\degr,-50\degr)$ is shown. The smoothing imposed was a Gaussian kernel with 
$\Delta\theta = 2\farcm0$ (FWHM). The shading indicates the value of the thermal Comptonisation $y$, which is proportional to 
$\mathrm{arsinh}(10^6\, y)$. The mesh size of the underlying Cartesian grid is $\sim14\arcsec$.}
\label{szobs_thszfield}
\end{figure}

\begin{figure}
\resizebox{\hsize}{!}{\includegraphics{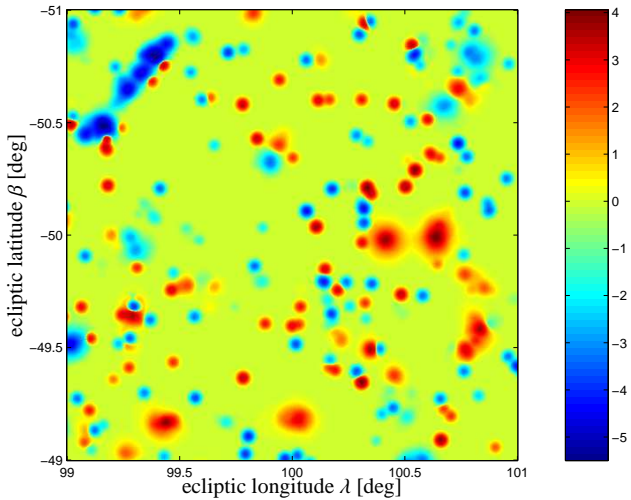}}
\caption{Detail of the kinetic Comptonisation map: A $2\degr\times2\degr$ wide cut-out is shown centered on the same position as 
Fig.~\ref{szobs_thszfield}, i.e. at the ecliptic coordinates $(\lambda,\beta) = (100\degr,-50\degr)$. The 
smoothing imposed was a Gaussian kernel with $\Delta\theta = 2\farcm0$ (FWHM). The kinetic Comptonisation $w$ is 
indicated by the shading, being proportional to $\mathrm{arsinh}(10^6\, w)$.}
\label{szobs_kinszfield}
\end{figure}

The sensitivity of \planck~is good enough to perform an SZ-survey of a significant part of the Hubble volume. Because the simulation of such a vast volume including hydrodynamics resolving scales as small as cluster substructure is beyond the capabilities of current computers, a hybrid approach has been pursued for the map construction: All-sky maps of the SZ-sky were constructed by using the light-cone output of the Hubble-volume simulation \citep{2001MNRAS.321..372J, 2000MNRAS.319..209C} as a cluster catalogue and template clusters from the small-scale gas-dynamical simulation \citep{2002ApJ...579...16W}. In this way, sky-maps were constructed which contain all clusters above $5\times 10^{13} M_{\sun}/h$ out to redshift $z=1.48$. The maps show the right 2-point halo correlation function, incorporate the evolution of the mass function and the correct distribution of angular sizes. Furthermore, they exhibit cluster substructure and deviations from the ideal cluster scaling relations induced e.g.\ by the departure from spherical symmetry. The velocities used for computing the kinetic SZ-effect correspond to the ambient density field. The map construction process and the properties of the resulting map are in detail described in \citet{2004_szmap}. Details of the thermal and kinetic SZ-maps can be seen in Figs.~\ref{szobs_thszfield} and \ref{szobs_kinszfield}, respectively.

\begin{table*}\vspace{-0.1cm}
\begin{center}
\begin{tabular}{lrrrrrrrrr}
\hline\hline
\vphantom{\Large A}%
\planck~ channel	& 1 & 2 & 3 & 4 & 5 & 6 & 7 & 8 & 9 \\
\hline
\vphantom{\Large A}%
centre frequency $\nu_0$				
& 30~GHz   & 44~GHz   & 70~GHz   & 100~GHz & 143~GHz & 217~GHz & 353~GHz & 545~GHz & 857~GHz 		\\
frequency window $\Delta\nu$			
& 3.0~GHz & 4.4~GHz & 7.0~GHz & 16.7~GHz & 23.8~GHz & 36.2~GHz & 58.8~GHz & 90.7~GHz & 142.8~GHz	\\
resolution $\Delta\theta$ (FWHM) & $33\farcm4$ & $26\farcm8$ & $13\farcm1$ & $9\farcm2$ &$7\farcm1$ & $5\farcm0$ & 
$5\farcm0$ & $5\farcm0$ & $5\farcm0$\\
noise level $\sigma_\mathrm{N}$	& 1.01~\mK & 0.49~\mK & 0.29~\mK & 5.67~\mK & 
4.89~\mK & 6.05~\mK & 6.80~\mK & 3.08~\mK & 4.49~\mK \\
\hline
thermal SZ-flux $\bra S_\mathcal{Y}\ket$	
& -12.2~Jy & -24.8~Jy & -53.6~Jy & -82.1~Jy & -88.8~Jy &  -0.7~Jy & 146.0~Jy & 76.8~Jy  & 5.4~Jy	\\
kinetic SZ-flux $\bra S_\mathcal{W}\ket$
& 6.2~Jy   & 13.1~Jy  & 30.6~Jy  &  55.0~Jy &  86.9~Jy & 110.0~Jy &  69.1~Jy & 15.0~Jy  & 0.5~Jy	\\
antenna temperature $\Delta T_\mathcal{Y}$ 	
& -440~\nK & -417~\nK & -356~\nK & -267~\nK & -141~\nK &  -0.5~\nK&   38~\nK &  8.4~\nK & 0.2~\nK	\\
antenna temperature $\Delta T_\mathcal{W}$ 	
& 226~\nK  & 220~\nK  & 204~\nK  &  179~\nK &  138~\nK &    76~\nK&   18~\nK &  1.6~\nK & 0.02~\nK	\\
\hline
\end{tabular}
\end{center}
\caption{Characteristics of \plancks \lfi-receivers (column 1-3) and \hfi-bolometers (column 4-9): centre frequency $\nu_0$, 
frequency window $\Delta\nu$ as defined in eqns.~(\ref{eqn_tlm_exp}) and (\ref{eq_freq_resp}), angular resolution 
$\Delta\theta$ stated in FWHM, effective noise level $\sigma_\mathrm{N}$, fluxes $\bra S_\mathcal{Y}\ket$ and $\bra 
S_\mathcal{W}\ket$ generated by the respective Comptonisation of $\mathcal{Y}=\mathcal{W}=1~\mathrm{arcmin}^2$ and the 
corresponding changes in antenna temperature $\Delta T_\mathcal{Y}$ and $\Delta T_\mathcal{W}$.}
\label{table_planck_channel}
\end{table*}

\section{Simulation of SZ-observations with {\em PLANCK}}\label{sect_sim}
In this section, the simulation is outlined: First, the foreground emission components considered are summarised 
(Sect.~\ref{sim_foreground}), and instrumental issues connected to sub-millimetric observations with \planck~are discussed (Sect.~\ref{sim_instrument}). The data products resulting from the simulation at this point will be spherical harmonics expansion coefficients $\bra S_{\ell m}\ket_\nu$ of the flux maps $S_{\nu}(\bmath{\theta})$ for all nine observing frequencies $\nu$, where the spectra have been convolved with \plancks frequency response windows and the spatial resolution of each channel is properly accounted for. Next, the signal extraction methodology based on matched and scale-adaptive filtering is described (Sect.~\ref{sim_filter}), followed by the application to simulated \planck-data (Sect.~\ref{sim_likelihood_synthesis}). The morphology of peaks in the filtered maps as a function of signal profile model parameters is discussed (Sect.~\ref{sim_likelihood_morphology}) and finally the algorithm for the extration of peaks in the filtered maps and the identification with objects in the cluster catalogue is described (Sect.~\ref{sim_peak_extraction}). A  description of the software tools and the foreground modelling used for our simulation can be found in \citet{2005astro.ph..8522R}.

\subsection{Foreground components}\label{sim_foreground}
We assume that the spectral properties of each emission component are isotropic, and that the amplitude of the emission is described by a suitably extrapolated template. It should be emphasised that this assumption is likely to be challenged, but the lack of observation makes it difficult to employ a more realistic scheme. Furthermore, it should be kept in mind that the extrapolation of the Galactic emission components by as much as three orders of magnitude in the case of synchrotron radiation is insecure. Details of the templates and the various emission laws are given in a precursor paper \citep{2004_szpsi}.

\begin{itemize}
\item{Cosmic microwave background: From the spectrum of $C_\ell$-coefficients generated with {\tt CMBfast} \citep{1996ApJ...469..437S} for the assumed \mbox{$\Lambda$CDM}-cosmology, a set of $a_{\ell m}$-coefficients was derived by using the {\tt synalm} code based on {\tt synfast} by \citet{1998elss.confE..47H}.}

\item{Galactic synchrotron emission: The modelling of the galactic synchrotron emission was based on an observation carried out by \citet{1981A&A...100..209H, 1982A&AS...47....1H} at an observing frequency of $\nu=408$~MHz, which has been adopted for \planck-usage by \citet{2002A&A...387...82G}. The spectral law for extrapolating the synchrotron flux incorporates a spectral break at $\nu=22$~GHz, which has been reported by the WMAP satellite.}

\item{Galactic dust emission: The thermal emission from Galactic dust \citep{1998ApJ...500..525S} used a two temperature model proposed by C. Baccigalupi, which yields a good approximation to the model introduced by D.J. Schlegel. The emission is modelled by a superposition of two Planck-laws with temperatures $T_1=9.4$~K and $T_2=16.2$~K at a fixed ratio.}

\item{Galactic free-free emission: Modelling of the Galactic free-free emission was based on an $H_\alpha$-template provided by \citet{2003ApJS..146..407F} and the spectral model proposed by \citet{1998PASA...15..111V}, which employs an approximate conversion from the $H_\alpha$ intensity to the free-free intensity parameterised with the plasma temperature and describes the spectral dependence of the free-free brightness temperature with a $\nu^{-2}$-law.}

\item{Emission from carbon monoxide in giant molecular clouds: Rotational transitions of carbon monoxide give rise to a series of lines at integer multiples of the frequency $\nu_\mathrm{CO}=115$~GHz. Its strength is given by the template provided by \citet{1996AAS...189.7004D,2001ApJ...547..792D}. The line strengths of higher harmonics of the transition were determined by assuming thermal equilibrium of the molecular rotational states with an ambient temperature of $T_\mathrm{CO}=20$~K.}

\item{Emission from planetary bodies of the solar system: We considered a model of the heat budget of the planetary surface with periodic heat loading from the Sun and internal sources of thermal energy, heat emission and conduction to the planet's atmosphere and the back-scattering onto the surface by the atmosphere. From the orbital elements of the planet as well as from the motion of the Lagrange point $L_2$ where \planck~will be positioned we derived the distance to the planet and its position in order to compute a sky map. In total, we considered 1200 asteroids apart from the five (outer) planets, excluding Pluto which is to faint to be detected by \planck.}
\end{itemize}

The list of foreground emission components which possibly hamper with the observation of Sunyaev-Zel'dovich clusters is still not complete. Microwave point sources such as AGN and starforming galaxies were not included, likewise the modelling of zodiacal light originating from interplanetary dust in the Solar system was not attempted. Concise descriptions of Galactic foregrounds at CMB frequencies derived from WMAP data are given by \citep{2003ApJS..148...97B} and \citet{2004astro.ph.10280P}.

\subsection{Instrumental issues}\label{sim_instrument}
The most important aspects related to the observations of the CMB sky by \planck~are the properties of the optical system, the noise introduced by the receivers and the frequency response, which will be summarised in this section:

\begin{itemize}
\item{\planck-beam shapes: The beam shapes of \planck~are approximated by azimuthally symmetric Gaussians $b(\theta) = 
\frac{1}{2\pi\sigma_\theta^2}\exp\left(-\frac{\theta^2}{2\sigma_\theta^2}\right)$ with $\sigma_\theta = 
\frac{\Delta\theta}{\sqrt{8\ln(2)}}$. The residuals from the idealised Gaussian shape are expected not to exceed the percent level. Table~\ref{table_planck_channel} gives the angular resolution $\Delta\theta$ for each channel.}

\item{Simulation of noise maps: Noise maps were generated by drawing Gaussian distributed random numbers from a distribution with zero mean and variance $\sigma_\mathrm{N}$ given by Table~\ref{table_planck_channel}. These numbers correspond to the noise for a single observation of a pixel by a single detector. Consequently, this number is downweighted by $\sqrt{n_\mathrm{det}}$  (assuming Poissonian statistics), where $n_\mathrm{det}$ denotes the number of redundant receivers per channel, because they provide independent surveys of the microwave sky. In a second step, exposure maps were derived by simulating scan paths with \planck~mission characteristics. Using the number of observations $n_\mathrm{obs}$ per pixel, it is possible to scale down the noise amplitudes by $\sqrt{n_\mathrm{obs}}$ and to obtain a realistic noise map for each channel.}

\item{Frequency response and superposition of emission components: Adopting the approximation of isotropy of the emission component's spectral behaviour, the steps in constructing spherical harmonics expansion coefficients $\bra S_{\ell m}\ket_{\nu_0}$ of the flux maps $S(\bmath{\theta},\nu)$ for all \planck~channels consist of deriving the expansion coefficients $a_{\ell m}$ of the template $a(\bmath{\theta})$,
\begin{equation}
a_{\ell m} = \int\dd\Omega\: a(\bmath{\theta})\, Y_\ell^m(\bmath{\theta})^*\leftrightarrow
a(\bmath{\theta}) = \sum_{\ell=0}^{\infty}\sum_{m=-\ell}^{+\ell} a_{\ell m}\, Y_\ell^m(\bmath{\theta})\mbox{,}
\label{eqn_ylm_decomp}
\end{equation}
converting the template amplitudes $a_{\ell m}$ to fluxes $S_{\ell m}$, extrapolate the fluxes with a known or assumed 
spectral emission law to \plancks observing frequencies, to finally convolve the spectrum with \plancks~frequency response 
window for computing the spherical harmonics expansion coefficients of the average measured flux $\bra S_{\ell m}\ket_{\nu_0}$ 
at nominal frequency $\nu_0$ by using:
\begin{equation}
\bra S_{\ell m}\ket_{\nu_0} 
= \frac{\int\dd\nu\: S_{\ell m}(\nu) R_{\nu_0}(\nu)}{\int\dd\nu\: R_{\nu_0}(\nu)}\mbox{.}
\label{eqn_tlm_exp}
\end{equation}
Here, $S_{\ell m}(\nu)$ describes the spectral dependence of the emission component considered, and $R_{\nu_0}(\nu)$ the 
frequency response of \plancks receivers centered on the fiducial frequency $\nu_0$. For all its channels, \plancks frequency response function $R_{\nu_0}(\nu)$ is well approximated by a top-hat function:
\begin{equation}
R_{\nu_0}(\nu) = 
\left\{
\begin{array}{l@{,\:}l}
1  & \nu\in\left[\nu_0-\Delta\nu,\nu_0+\Delta\nu\right] \\
0  & \nu\notin\left[\nu_0-\Delta\nu,\nu_0+\Delta\nu\right] 
\end{array}
\right.
\label{eq_freq_resp}
\end{equation}
The centre frequencies $\nu_0$ and frequency windows $\Delta\nu$ for \plancks receivers are summarised in 
Table.~\ref{table_planck_channel}. In the final step, the averaged fluxes $\bra S_{\ell m}\ket_{\nu_0}$ for each emission component are added to yield the expansion coefficients of the flux map.}
\end{itemize}

\subsection{Construction of optimised filter kernels}\label{sim_filter}
The basis of our signal extraction method is the concept of matched and scale-adaptive filtering pioneered by \citet{2001ApJ...552..484S} and \citet{2002MNRAS.336.1057H}, which we generalised to the case of spherical data sets and spherical harmonics $Y_{\ell m}(\bmath{\theta})$ as the harmonic system replacing Fourier transforms and plane waves of the case of a flat geometry. The theory of matched and scale-adaptive filtering of multifrequency data sets is beautifully developed by the above mentioned authors in their formulation as the solution to a functional variation problem with increasing complexity. Filter kernels $\psi(\left|\bmath{\theta}\right|, R)$ constructed for finding objects of a certain scale $R$ modify the sky map  $w(\bmath{\theta})$ by convolution and are required to minimise the variance $\sigma^2_w(R)$ of the filtered map $w(\bmath{\theta},R)$ while fulfilling certain conditions: 


\begin{enumerate}
\item{There should exist a scale $R_0$ for which the value of the filtered field $\bra w(R_0)\ket$ at the position of a source is maximal,}
\item{the filter should be unbiased, i.e. $\bra w(R_0)\ket$ at the position of a source should be proportional to the amplitude of the underlying signal, and}
\item{the variance $\sigma_w^2(R)$ should have a minimum at this scale $R_0$.}

\end{enumerate}
In short, the filtered field should yield maximised signal-to-noise values for the peaks, while being linear in the signal and allowing the measurement of sizes of objects. We restrict ourselves to spherically symmetric source profiles superimposed on a fluctuating background which is a realisation of a homogeneous and isotropic Gaussian random field characterised by its power spectrum. The filter resulting from the variation with the boundary conditions (i) and (ii) is called matched filter, and the filter obeying all three conditions is refered to as the scale-adaptive filter.

Solving this variational problem in its extension to multifrequency data and spherical maps yields spherical harmonics expansion coefficients $\psi_\nu(\ell)$ for the filter kernels of each observational channel $\nu$ as a function of the assumed profile of the source, of the spectral variation of the source flux with frequency and of the auto- and cross correlation power spectra $C_{\nu_i\nu_j}(\ell)$. The relative normalisation of the filter kernels in different channels encodes the optimised linear combination coefficients for co-adding the filtered maps. Apart from that, one obtains an analytical expression for the fluctuation amplitude $\sigma^2=\sum_\ell\sum_{\nu_i}\sum_{\nu_j}\psi_{\nu_i}(\ell)C_{\nu_i\nu_j}(\ell)\psi_{\nu_j}(\ell)$ of the filtered and co-added maps, such that any peak height can be expressed in units of the map's standard deviation, which directly allows to quantify whether a certain peak is likely to be a genuine signal or a mere background fluctuation.

Filter kernels following from the matched and scale-adaptive multifrequency filtering algorithm were subjected to a thorough analysis. They are tested on two different data sets, one containing just CMB fluctuations and (non-isotropic) instrumental noise, and a second data set, which comprises all foregrounds in addition. From the comparison of the two data sets one will be able to quantify by how much the number of detections drop due to the foreground components and how uniform the cluster distribution will be provided the removal of foregrounds can be done in efficiently. Details of the cross-channel correlation properties of the foregrounds as well as of the noise and the derivation of filter kernels and their properties are discussed in a precursing paper \citep{2004_szpsi}.

\subsection{Filter construction and synthesis of likelihood maps}\label{sim_likelihood_synthesis}
Filter kernels optimised for detecting a family of King-profiles $y(\theta)\propto\left[1+(\theta/\theta_c)^2\right]^{-\lambda}$ were derived for a range of core radii $\theta_c$ and asymptotic slopes $\lambda$. Specifically, seven values of $\theta_c$,
\begin{equation}
\theta_c = 0\farcm0,\quad 1\farcm0,\quad 2\farcm0,\quad4\farcm0,\quad8\farcm0,\quad16\farcm0,\quad32\farcm0,
\end{equation}
and five values of $\lambda$,
\begin{equation}
\lambda = 0.6,\quad0.8,\quad1.0,\quad1.2,\quad1.4
\end{equation} 
were considered, keeping the large range in core radii in mind. Using different values for $\lambda$ is motivated by deviations from the generic baryonic profile and by asymmetric clusters. The sky maps were convolved with the filter kernels, co-added, normalised to unit variance, as described in the previous section and synthesised to yield likelihood maps. In the synthesis, all multipole coefficients up to $\ell=4096$ have been considered and the angular resolution of the resulting maps ($N_\mathrm{side}=1024$, pixel side length $\simeq3\farcm4$) is high enough to resolve single likelihood peaks.

An important numerical issue of spherical harmonic transforms is the fact that the variance (measured in real space) of a map synthesised from the $a_{\ell m}$-coefficients is systematically smaller with increasing $\ell$ than the variance $C(\ell)$ required by the $a_{\ell m}$-coefficients on the scale $\Delta\theta\simeq\pi/\ell$. This is compensated by an empirical function, the so-called {\em pixel window}, which lifts the amplitudes $a_{\ell m}$ towards increasing values of $\ell$ prior to the reconstruction. This effectively results in higher signal-to-noise ratios of the detected clusters. In the numerical derivation of filter kernels very low multipoles below $\ell\leq3$ were excluded because of numerical instabilities in a matrix inversion and set artificially to zero. An important consequence of this will be discussed in Sect.~\ref{result_spatial_distrib}.

\subsection{Morphology of SZ-clusters in filtered sky-maps}\label{sim_likelihood_morphology}
\begin{figure*}
\begin{center}
\begin{tabular}{cccc}
\resizebox{4cm}{!}{\includegraphics{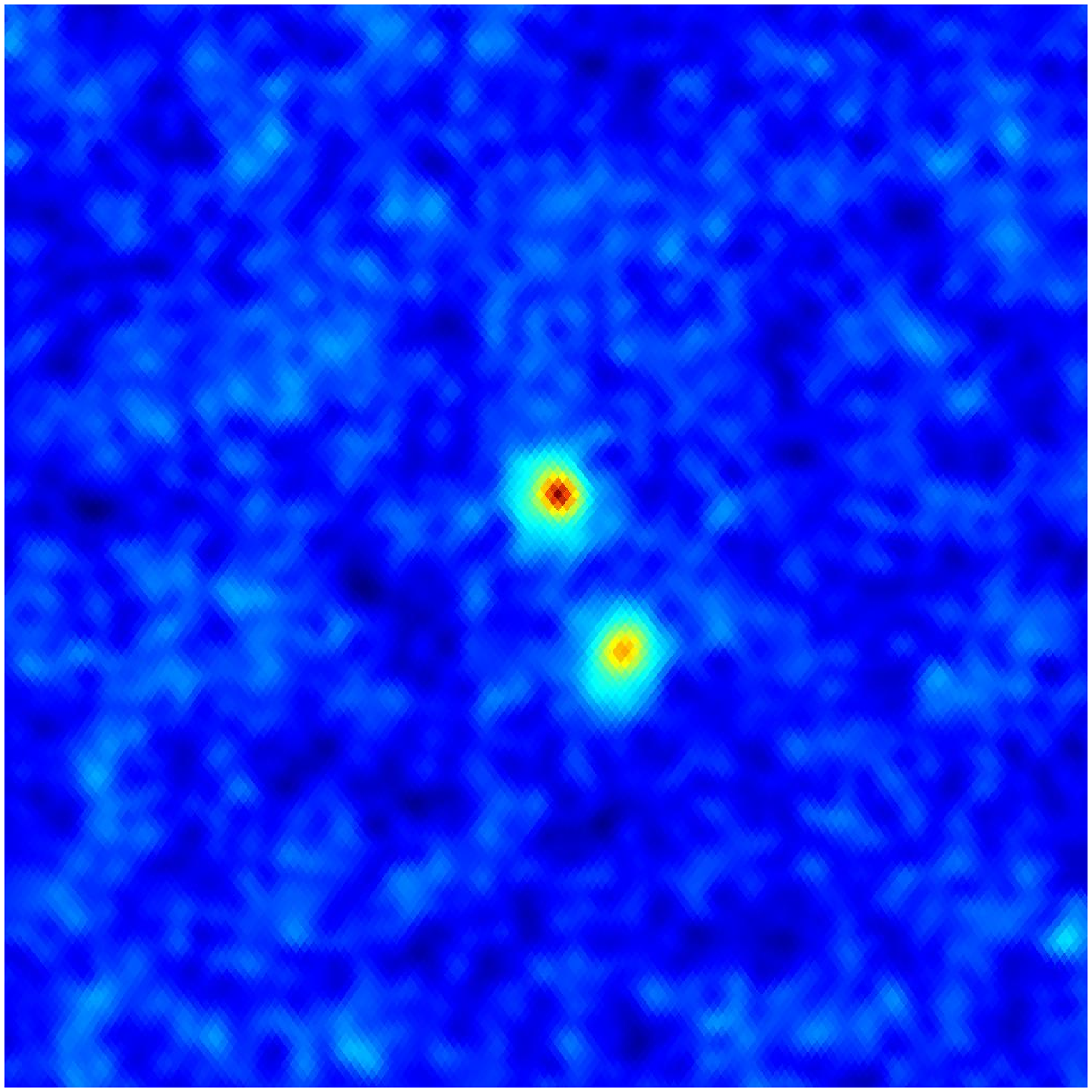}} &
\resizebox{4cm}{!}{\includegraphics{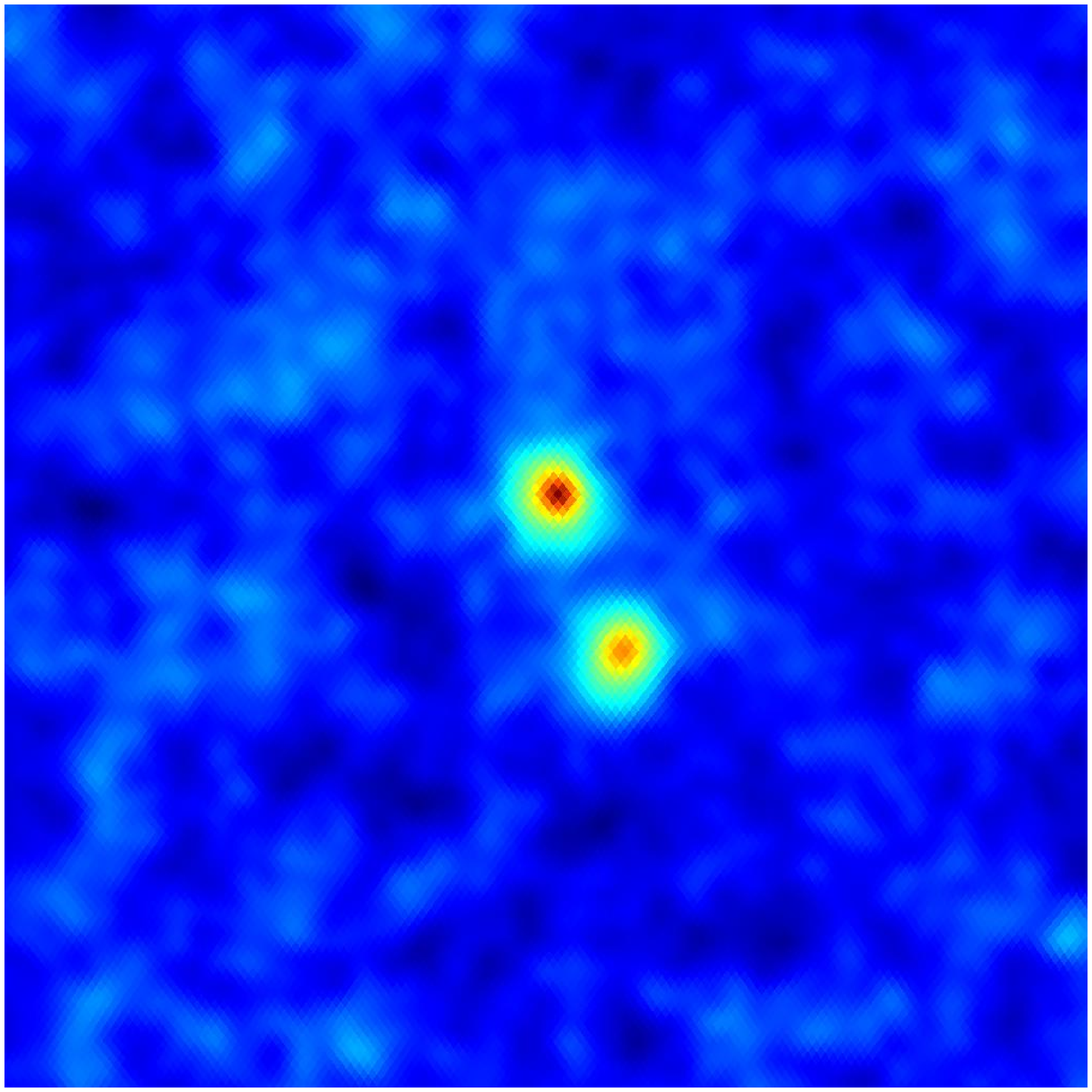}} &
\resizebox{4cm}{!}{\includegraphics{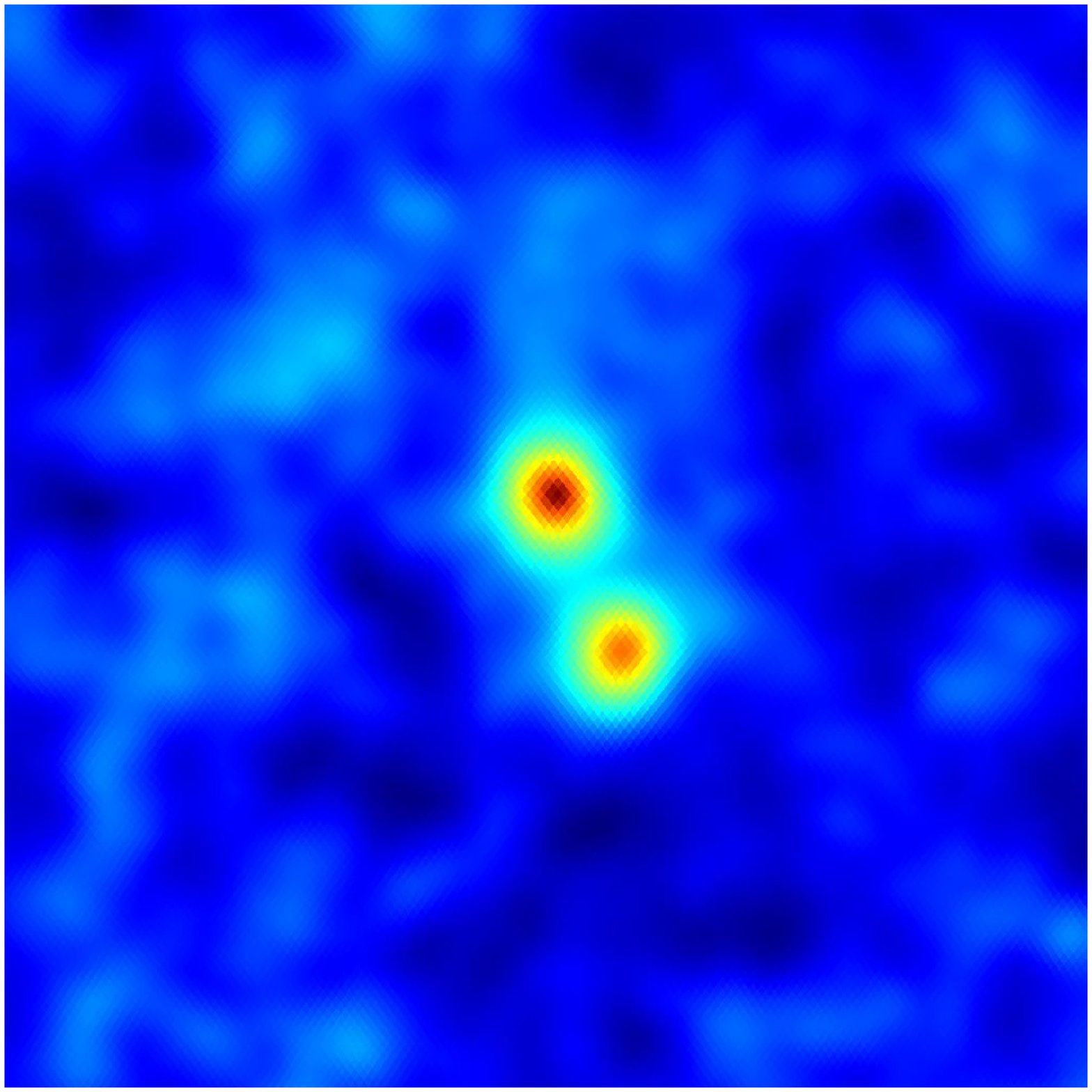}} &
\resizebox{4cm}{!}{\includegraphics{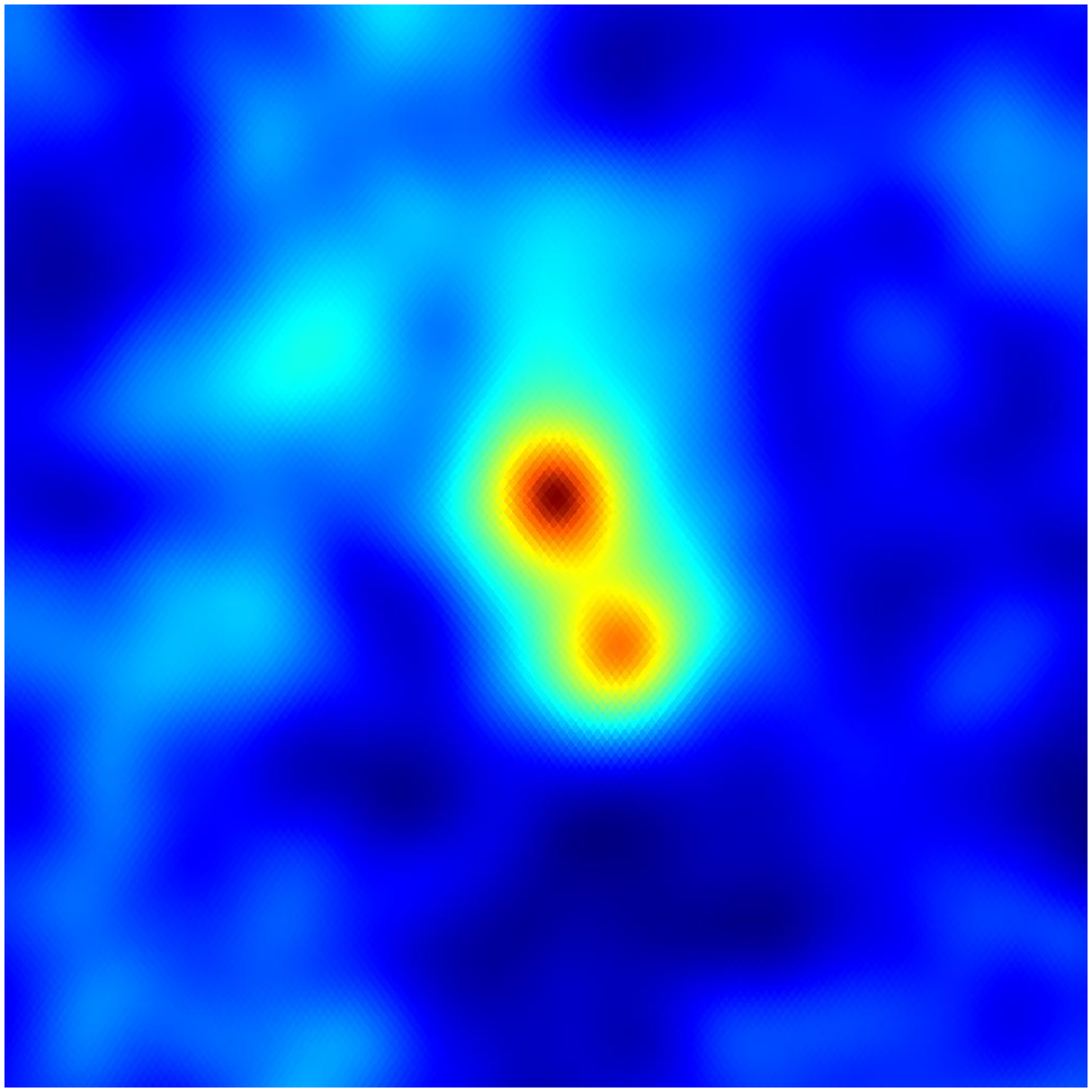}} \\
$\theta_c=4\farcm0$ $\rightarrow$ $18.93\sigma$ &
$\theta_c=8\farcm0$ $\rightarrow$ $17.86\sigma$ &
$\theta_c=16\farcm0$ $\rightarrow$ $14.17\sigma$ &
$\theta_c=32\farcm0$ $\rightarrow$ $9.37\sigma$ \\
\resizebox{4cm}{!}{\includegraphics{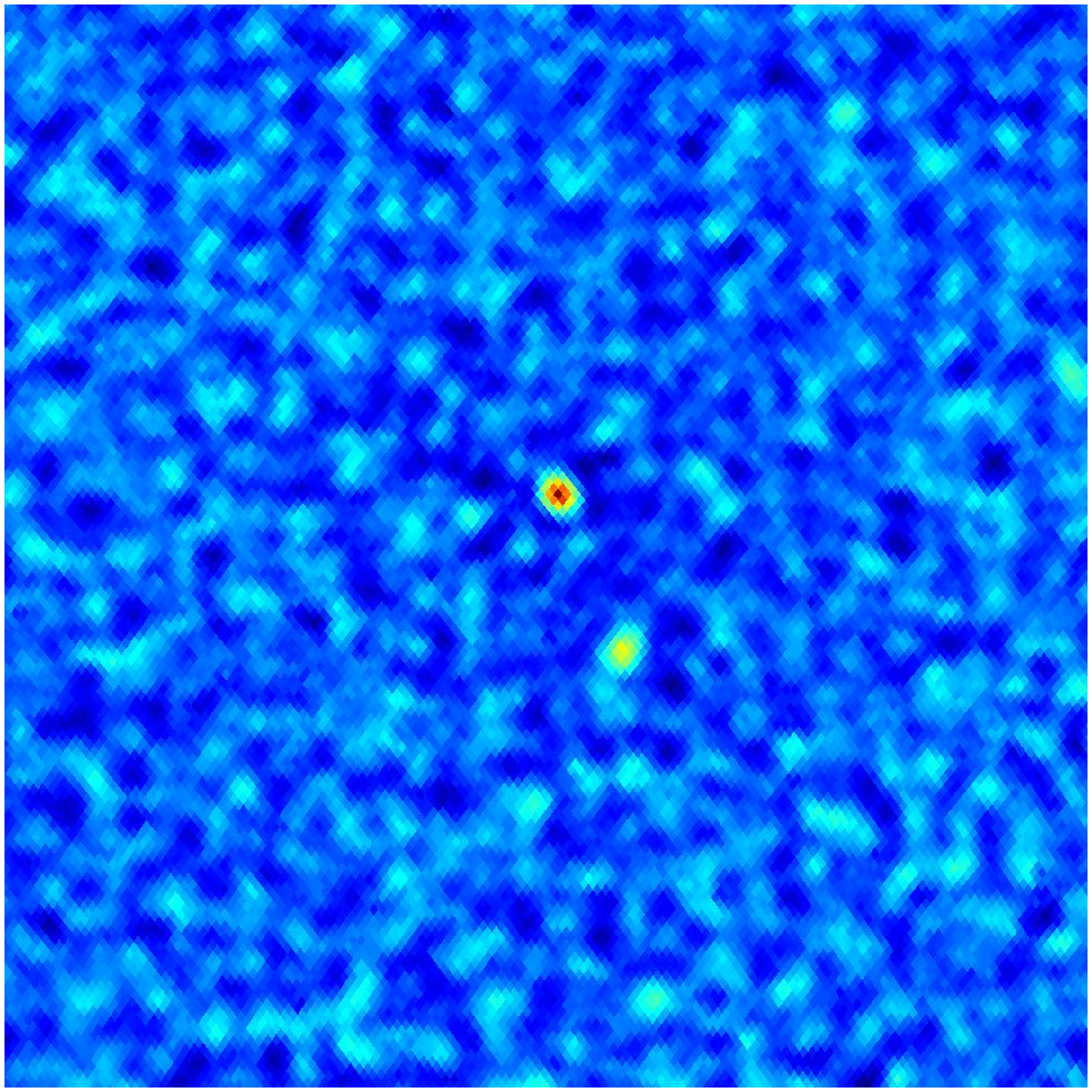}} &
\resizebox{4cm}{!}{\includegraphics{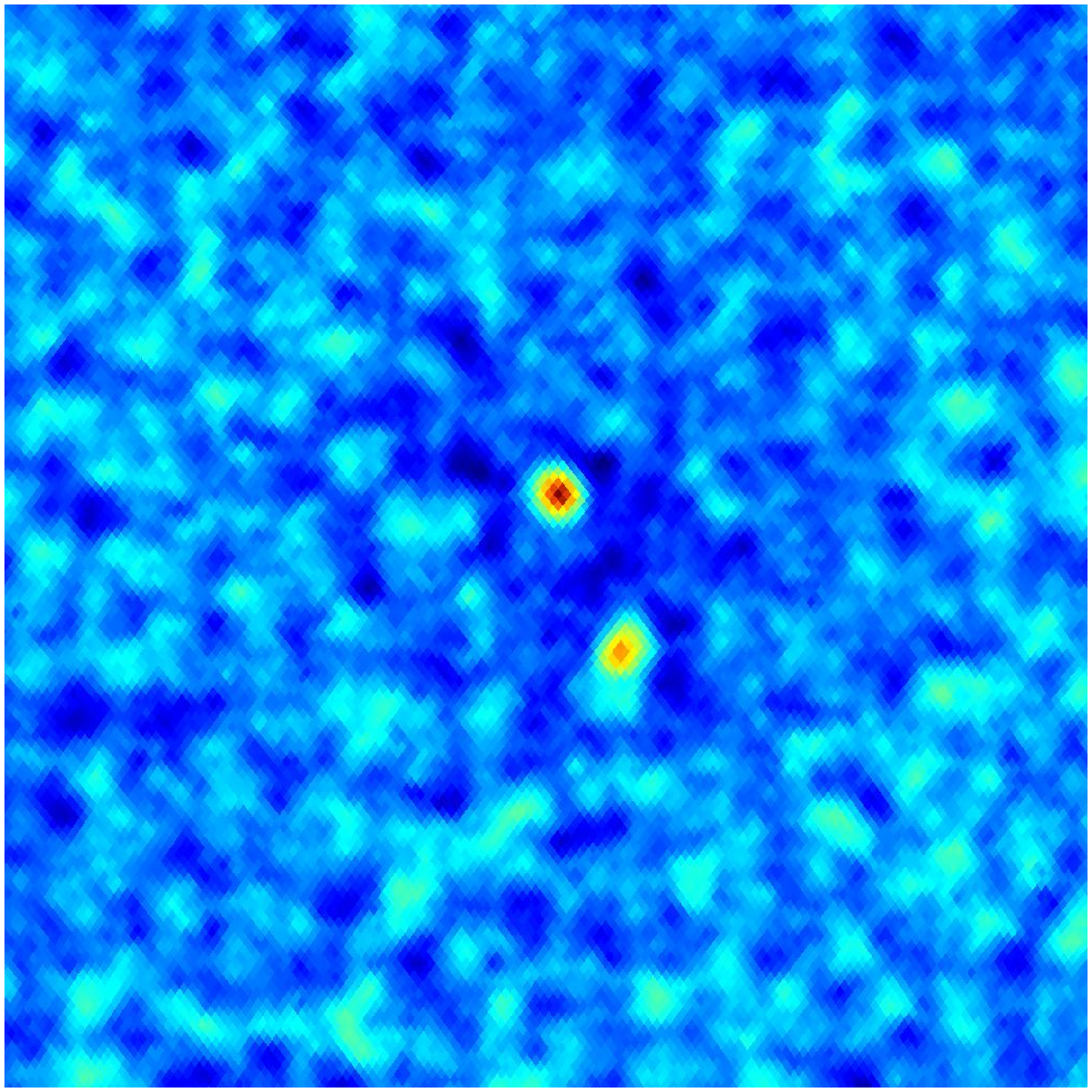}} &
\resizebox{4cm}{!}{\includegraphics{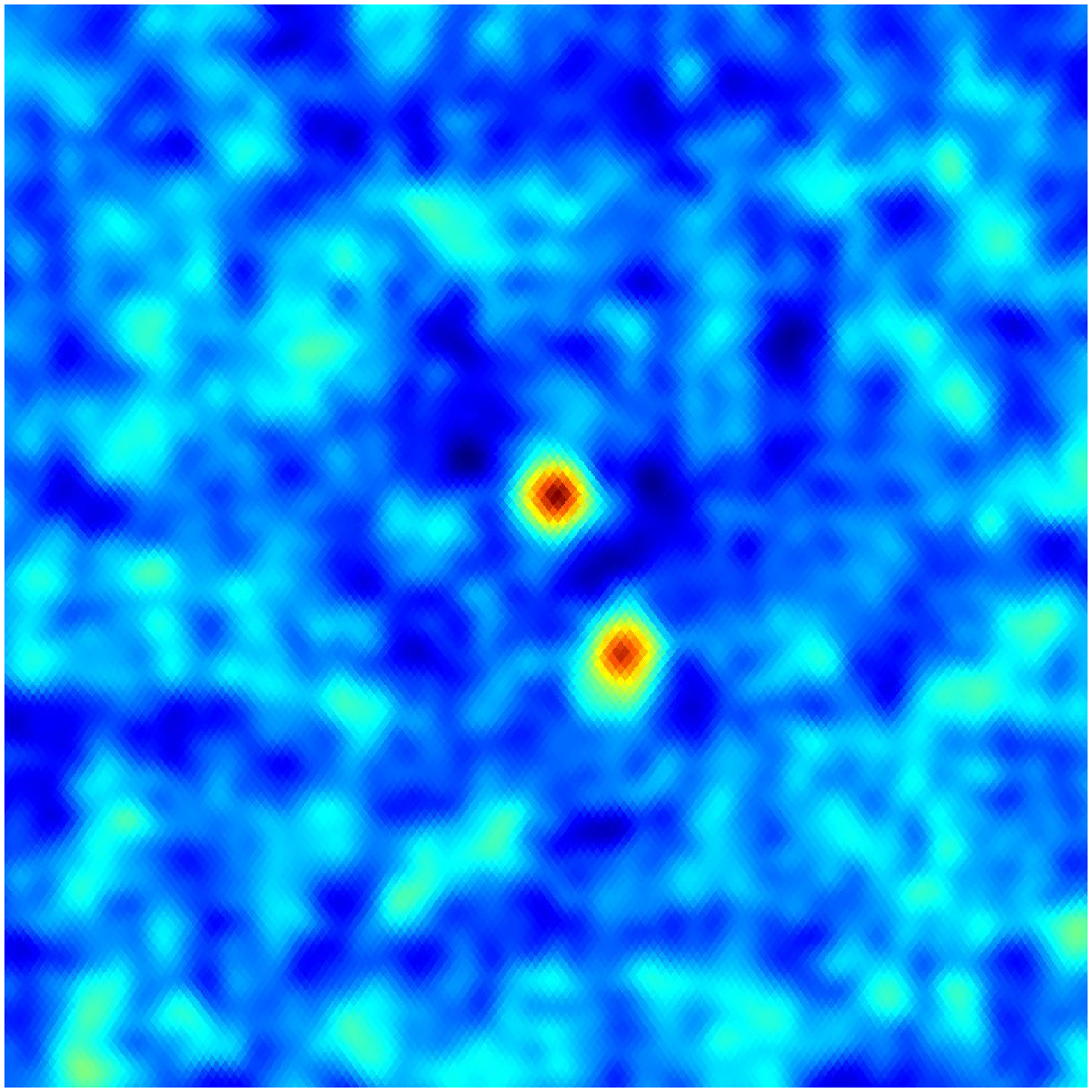}} &
\resizebox{4cm}{!}{\includegraphics{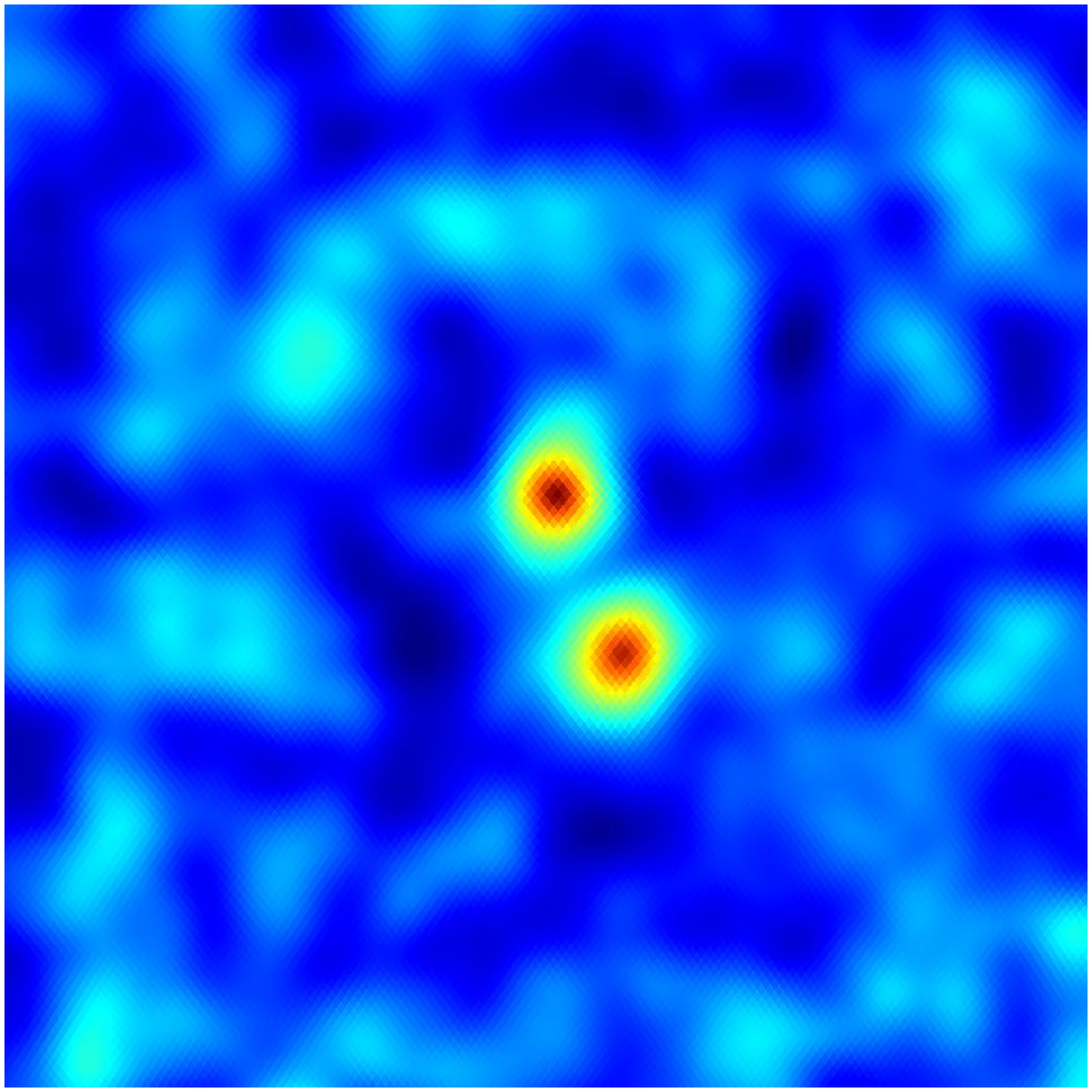}} \\
$\theta_c=4\farcm0$ $\rightarrow$ $11.10\sigma$ &
$\theta_c=8\farcm0$ $\rightarrow$ $10.15\sigma$ &
$\theta_c=16\farcm0$ $\rightarrow$ $9.33\sigma$ &
$\theta_c=32\farcm0$ $\rightarrow$ $9.74\sigma$ \\
\end{tabular}
\end{center}
\caption{An association of two clusters at $z\simeq0.1$, extracted with the matched multifilter (top row) and with the scale-adaptive multifilter (bottom row) from a map containing all Galactic compontents, CMB fluctuations and instrumental noise. The panel gives the likelihood maps and the statistical significances of the detection of the cluster at the image centre in units of $\sigma$, for varying values of $\theta_c$ with $\lambda$ fixed at $\lambda=1.0$. The side length of the panels is $4\degr$.}
\label{szobs_fig_phi_comparison}
\end{figure*}

Fig.~\ref{szobs_fig_phi_comparison} gives an impression how the morphology of a peak in the likelihood map changes if filter kernels optimised for the detection of profiles with varying diameter and asymptotic behaviour are used. We picked an association of two clusters at a redshift of $z\simeq0.1$, which generates a signal strong enough to yield a significant detection irrespective of the choice of $\theta_c$ and $\lambda$. 

The matched filter yields larger values for the detection significance, which is defined to be the signal-to-noise ratio of the central object, in comparison to the scale-adaptive filter for that particular pair of clusters. Secondly, if filters optimised for large objects, i.e. large $\theta_c$ and small $\lambda$ are used, the two peaks merge in the case of the matched filter, but stay separated in case of the scale-adaptive filter. Hence the scale-adaptive filter is more appropriate in the investigation of neighbouring objects. Additionally, the matched filter seems to be more sensitive to the choice of $\theta_c$ and $\lambda$. Within the range of these two parameters considered here, the significance of the cluster detection under consideration varies by a factor of four in the case of the matched filter, but changes only by 25\% in the case of the scale-adaptive filter.

\subsection{Peak extraction and cluster identification}\label{sim_peak_extraction}
It is an important point to notice that cluster positions derived from \planck~are not very accurate. In this analysis, the SZ-clusters are extended themselves and possibly asymmetric, they are convolved with \planck's instrumental beams in the observation and reconstructed from filtered data, where an additional convolution with a kernel is carried out. Furthermore, the pixelisation is relatively coarse (typically a few arcmin). All these effects add up to a position uncertainty of a few tens of arc minutes, depending on the filter kernel.

All peaks above $3\sigma$ were extracted from the synthesised likelihood maps and cross checked with a cluster catalogue. A peak was taken to be a detection of a cluster if its position did not deviate more than $30\farcm0$ from the nominal cluster position. Peaks that did not have a counterpart with integrated Comptonisation $\mathcal{Y}$ larger than a predefined threshold value were registered as false detections, likewise peaks were not considered that did not exceed the threshold value of $3\sigma$ in more than two contiguous pixels. In this way, a catalogue is obtained which is essentially free of false detections and where the fraction of unidentified peaks amounts to $5-7$\% for a realistic threshold of $\mathcal{Y}_\mathrm{min}=3\times10^{-4}~\mathrm{arcmin}^2$ \citep{1997mba..proc..413H, 2001A&A...370..754B}. The cluster catalogues following from observations with specific $(\theta_c,\lambda)$-pairs of parameters were merged to yield summary catalogues for both filter algorithms and both noise compositions. If more than one cluster is found in the aperture, the cluster with the largest value for the integrated Comptonisation is assumed to generate the signal. In the merging process, we determine which choice of $(\theta_c,\lambda)$ yielded the most significant detection for a given object.

\section{Results}\label{sect_result}
First of all, we investigate the noise properties of the likelihood maps (Sect.~\ref{result_noise}), followed by an analysis of the detection significances (Sect.~\ref{result_significance}) the different filter algorithms are able to yield. Then, the number of detected clusters as a function of model profile parameters is investigated (Sect.~\ref{result_paramspace}). The population of SZ-clusters in the mass-redshift plane (Sect.~\ref{result_mzplane}), the distribution of the position accuracies (Sect.~\ref{result_sigma}) and the spatial distribution of clusters (Sect.~\ref{result_spatial_distrib}) are the main results of this work. Finally, the detectability of the kinematic SZ-effect is addressed (Sect.~\ref{result_kinsz}).

\subsection{Noise in the filtered and co-added maps}\label{result_noise}
\begin{figure}
\resizebox{\hsize}{!}{\includegraphics{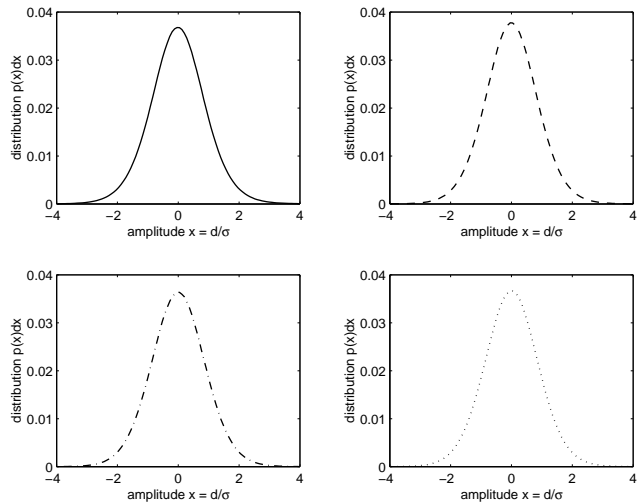}}
\caption{Distribution of pixel amplitudes $d$ of the filtered and co-added maps, normalised to the variance $\sigma$ predicted in the filter kernel derivation, for a data set including CMB fluctuations and instrumental noise, filtered with the matched filter (upper left, solid line), for a data set including Galactic foregrounds in addition (upper right, dashed line), for a data set containing the CMB and instrumental noise, filtered with the scale-adaptive filter (lower left, dash-dotted line) and finally a data set with CMB, instrumental noise and Galactic foregrounds, filtered with the scale-adaptive filter (lower right, dotted line). The filters have been optimised for the detection of beam-shaped profiles.}
\label{szobs_filter_sigma}
\end{figure}

In this section, the statistical properties of the noise in the filtered maps is examined. The filter construction algorithm gives the variance $\sigma$ of the filtered and co-added fields as a function of filter shapes 
$\psi_{\nu_i}(\ell)$ and cross-channel power spectra $C_{\nu_i\nu_j}(\ell)$ by virtue of $\sigma^2=\sum_\ell \sum_i 
\sum_j \psi_{\nu_i}(\ell) C_{\nu_i\nu_j}(\ell) \psi_{\nu_j}(\ell)$. Due to deviations from Gaussianity of many noise components 
considered (especially Galactic foregrounds), it is important to verify if the variance is still a sensible number. 
Fig.~\ref{szobs_filter_sigma} gives the distribution of pixel amplitudes for a combination of noise components and filtering 
schemes. 

Although the distribution of pixel amplitudes seems to follow a Gaussian distribution with zero mean and unit variance in all cases, there are notable deviations from this first impression. As summarised in Table~\ref{szobs_table_noise}, the mean of the distributions is compatible with zero in all cases, but the standard deviation is less than unity. Furthermore, the kurtosis of all distributions is nonzero, hence they are more outlier-prone as the normal distribution (barykurtic), which leads to a misestimation of statistical significances of peaks based on the assumption of unit variance of the filtered map, which the filtered map should have due to the renormalisation. This effect is strongest in the case of the matched filter. For the derivation of these numbers, only pixels with amplitudes smaller than $\left|d\right|\leq4\sigma$ have been considered, such that the statistical quantities are dominated by the noise to be examined and not by the actual signal. The distributions are slightly skewed towards positive values, which is caused by weak signals below $4\sigma$. The near-Gaussianity suggests that the residual noise in the filtered map is mostly caused by uncorrelated pixel noise and filters seem to be well capable of suppressing unwanted foregrounds.

Is it important to notice that the comparatively low threshold of $3\sigma$ imposed for extracting the peaks alone would yield a considerable number of false detections. This motivated the rather complicated algorithm outlined in Sect.~\ref{sim_peak_extraction}. Supposing that the variance of the filtered maps is mainly caused by uncorrelated pixel noise which is smoothed to an angular scale of $\simeq20^\prime$ by the instrumental beam and by the filters causes the filtered maps to be composed of $4\pi(180/\pi)^2\cdot3^2\simeq4\times10^5$ unconnected patches. Of these patches, a fraction of $\mathrm{erfc}(3/\sqrt{2})\simeq10^{-4}$ naturally fluctuates above the threshold of $3\sigma$. In this way a total number of $\simeq400$ patches have significances above $3\sigma$. The requirement that the counterpart of the peak in the cluster catalogue generates a Comptonisation above a (conservative) value of $\mathcal{Y}_\mathrm{min}$, i.e. that a cluster candidate is confirmed by spectroscopy, removes these false peaks from the data sample.

\begin{table*}
\vspace{-0.1cm}
\begin{center}
\begin{tabular}{llcccc}\hline\hline
\vphantom{\Large A}%
filter algorithm & data set & mean $\mu$ & variance $\sigma$ & skewness $s$ & kurtosis $k-3$ \\
\hline
\vphantom{\Large A}%
matched		& CMB + noise		& $-0.0038\pm0.0005$ & $0.9272\pm0.0003$ & $0.0334$ & $0.5297$ \\
matched		& CMB + noise + Galaxy	& $-0.0009\pm0.0005$ & $0.8902\pm0.0003$ & $0.0154$ & $0.4232$ \\
scale-adaptive	& CMB + noise		& $-0.0012\pm0.0005$ & $0.9090\pm0.0004$ & $0.0142$ & $0.2923$ \\
scale-adaptive	& CMB + noise + Galaxy	& $-0.0005\pm0.0005$ & $0.9023\pm0.0004$ & $0.0076$ & $0.3125$ \\
\hline
\end{tabular}
\end{center}
\caption{Statistical properties of the filtered and co-added maps, derived from the first four moments of the amplitude distributions in Fig.~\ref{szobs_filter_sigma}, for all data sets and filter algorithms. The filters have been optimised for the detection of beam-shaped profiles. The errors given for the mean $\mu$ and standard deviation $\sigma$ of the distribution of pixel amplitudes correspond to 95\% confidence intervals.}
\label{szobs_table_noise}
\end{table*}

\subsection{Detection significances and total number of detections}\label{result_significance}
\begin{figure}
\resizebox{\hsize}{!}{\includegraphics{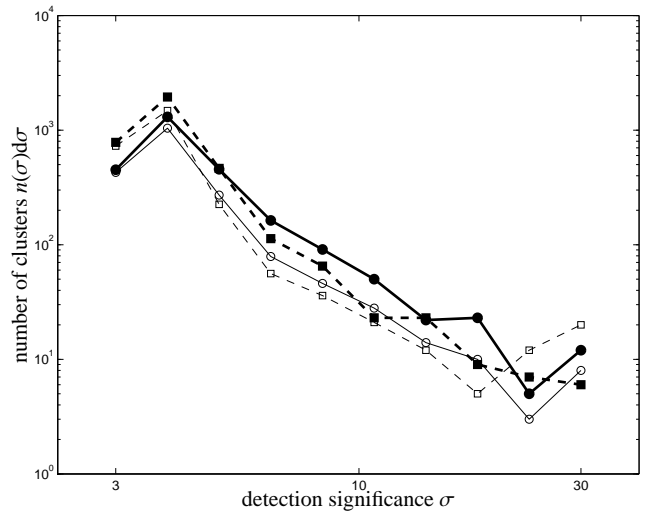}}
\caption{Distribution $n(\sigma)\dd\sigma$ of the detection significances $\sigma$, for the matched filter (solid 
line, circles) in comparison to the scale-adaptive filter (dashed line, squares). The distributions are given for the clean 
data set including only the CMB, both SZ-effects and instrumental noise (thick lines, closed symbols) and in comparison, the data set where all Galactic foreground components are included in addition (thin lines, open symbols).}
\label{szobs_sigmadistro}
\end{figure}

The distribution of detection sigificances is given in Fig.~\ref{szobs_sigmadistro}. One obtains about $10^3$ detections at the significance threshold which drops to a few highly significant detections exceeding $20\sigma$. At small $\sigma$, the scale-adaptive filter yields more detections than the matched filter, which catches up at roughtly $5\sigma$. 

The total number of detections for each filter algorithm, for each data set and for two values of the minimally required Comptonisation $\mathcal{Y}_\mathrm{min}$ for spectroscopic confirmation are compiled in Table~\ref{szobs_table_detect}. Due to its better yield of detections marginally above the threshold the scale-adaptive filter outperforms the matched filter by almost 30\%. The reason for the increased number of low-significance detections is the systematically higher value of the variance of the residual noise field in the case of the scale-adaptive filter. The number of detections decreases by $\simeq25$\% if Galactic foregrounds are included, relative to the data set containing only CMB fluctuations and instrumental noise. In a realistic observation, one can expect a total number of $\sim6\times10^3$ clusters of galaxies, compared to $\simeq8\times10^3$ clusters if only the CMB and instrumental noise were present. When comparing the total number of detections to analytic estimates \citep[e.g. ][]{1997A&A...325....9A, 2001MNRAS.325..835K, 2001A&A...370..754B}, it is found that the number of clusters detected here is smaller, by a factor of less than two. 

One should keep in mind that the noise due to PLANCK's scanning paths is highly structured on the cluster scale and below, such that the assumption of isotropy of the noise is not valid. This has two important consequences: Firstly, assuming a simple flux threshold in analytic estimates is not valid because the noise is not uniform on the cluster scale and secondly the assumption of isotropy which is essential to the filter construction is violated which affects the sensitivity of the filters. 

\begin{table*}
\vspace{-0.1cm}
\begin{center}
\begin{tabular}{llcc}\hline\hline
\vphantom{\Large A}%
filter algorithm & data set & $\mathcal{Y}_\mathrm{min}=10^{-3}~\mathrm{arcmin}^2$ &$\mathcal{Y}_\mathrm{min}=3\times10^{-4}~\mathrm{arcmin}^2$\\
\hline
\vphantom{\Large A}%
matched filter		& CMB + noise		&2402	&5376	\\
matched filter		& CMB + noise + Galaxy	&1801	&4199	\\
scale-adaptive filter	& CMB + noise		&3234	&8020	\\
scale-adaptive filter	& CMB + noise + Galaxy	&2428	&6270	\\
\hline
\end{tabular}
\end{center}
\caption{Total number of detections in both data sets and with both filters, and for two values for minimally required Comptonisation.}
\label{szobs_table_detect}
\end{table*}

\subsection{Cluster detectability as a function of filter parameters}\label{result_paramspace}
The way the significance of a detection of a cluster changes when the core size $\theta_c$ and the asymptotic slope $\lambda$ are varied is illustrated in Fig.~\ref{szobs_fig_phi_comparison} for the matched filter. In general, the matched filter yields significances that are almost twice as large in comparison to the scale-adaptive filter for the specific example considered and consequently finds more clusters above a certain detection threshold. Furthermore, the matched filter shows a stronger dependence of the significance on the filter parameters $\theta_c$ and $\lambda$: The significance for the detection of the same object varies by a factor of four in case of the matched filter but only by 25\% in the case of the scale-adaptive filter. This means that the derivation of cluster properties based on the filter parameter that yielded the most significant detection is likely to work for the matched filter, but not for the scale-adaptive filter. It should be emphasised, however, that the scale-adaptive filter keeps the likelihood distributions of the two objects from merging, in contrast to the matched filter. For that reason, the scale-adaptive filter may be better suited for the investigation of associations and pairs of SZ-clusters.

\begin{figure}
\resizebox{\hsize}{!}{\includegraphics{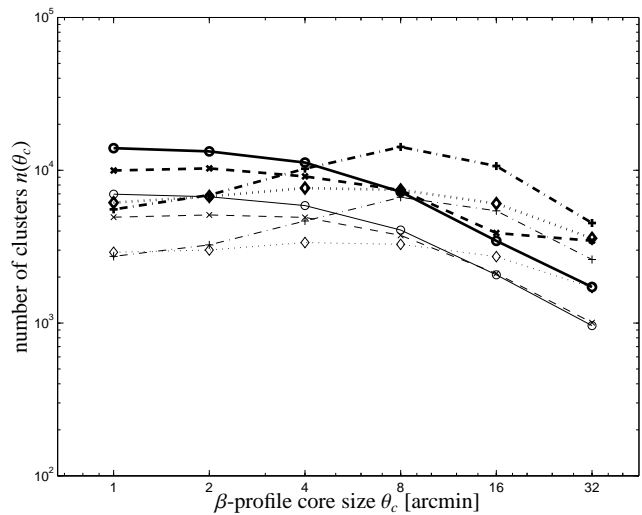}}
\caption{Number density $n(\theta_c)$ of clusters as a function of the filter parameter core size $\theta_c$, for a data set 
including CMB fluctuations and instrumental noise, filtered with the matched filter (circles, solid line), for a data set 
including Galactic foregrounds in addition (crosses, dashed line), for a data set containing the CMB and instrumental noise, 
filtered with the scale-adaptive filter (plus signs, dash-dotted line) and finally a data set with CMB, instrumental noise and 
Galactic foregrounds, filtered with the scale-adaptive filter (diamonds, dotted line). The thick and thin lines denote detections 
and peaks above $10^{-3}~\mathrm{arcmin}^2$ and $3\times10^{-4}~\mathrm{arcmin}^2$, respectively.}
\label{szobs_num_filter_theta}
\end{figure}

Fig.~\ref{szobs_num_filter_theta} shows the number density of detectable clusters as a function of the King-profile's core 
size $\theta_c$ that entered the filter construction. Whereas the matched filter yields most detections at small values of $\theta_c$, the scale-adaptive filter is better suited to detect extended objects. Most of the detections are registered at core sizes $\theta_c=8^\prime$. Additionally, the scale-adaptive filter's capability of detecting extended objects suffers from the inclusion of Galactic foregrounds, which cause the total number of detections to drop by 20\%. In contrast, the matched filter is able to deliver a comparable performance for all values of $\theta_c$ if Galactic foregrounds are included.

\begin{figure}
\resizebox{\hsize}{!}{\includegraphics{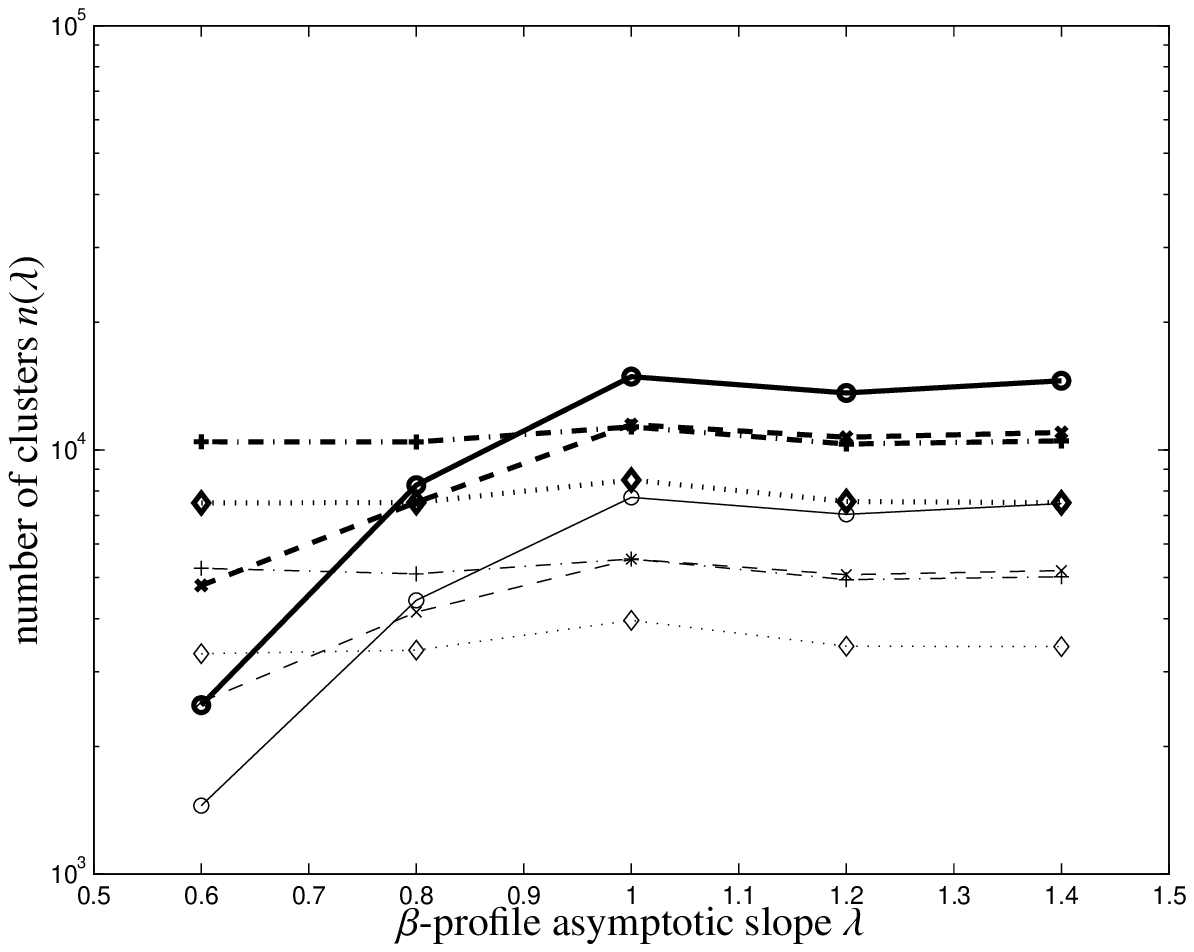}}
\caption{Number density $n(\lambda)$ of clusters as a function of the filter parameter asymptotic slope $\lambda$, for a data set including CMB fluctuations and instrumental noise, filtered with the matched filter (circles, solid line), for a data set 
including Galactic foregrounds in addition (crosses, dashed line), for a data set containing the CMB and instrumental noise, 
filtered with the scale-adaptive filter (plus signs, dash-dotted line) and finally a data set with CMB, instrumental noise and 
Galactic foregrounds, filtered with the scale-adaptive filter (diamonds, dotted line). The thick and thin lines denote detections  and peaks above $10^{-3}~\mathrm{arcmin}^2$ and $3\times10^{-4}~\mathrm{arcmin}^2$, respectively.}
\label{szobs_num_filter_lambda}
\end{figure}

\begin{figure*}
\begin{center}
\begin{tabular}{cc}
\resizebox{0.47\linewidth}{!}{\includegraphics{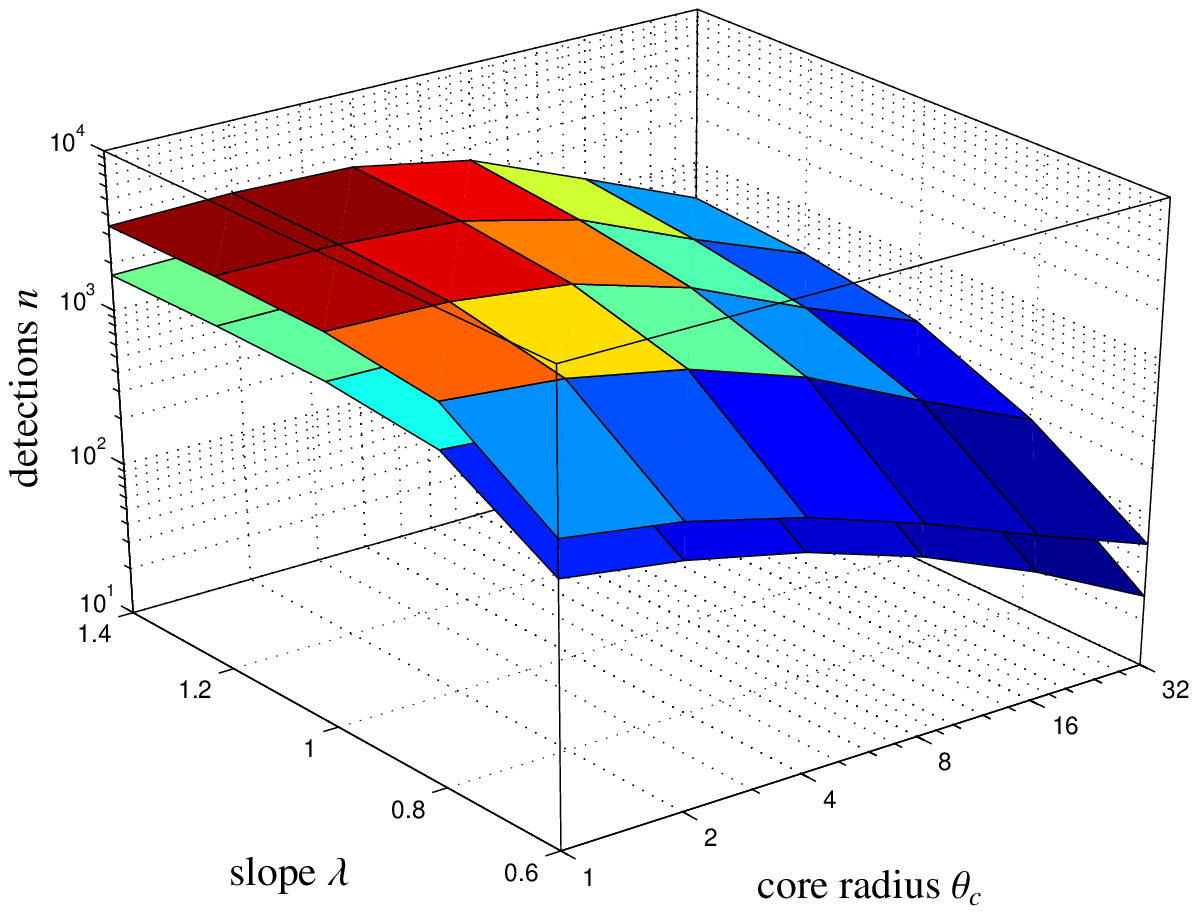}} & 
\resizebox{0.47\linewidth}{!}{\includegraphics{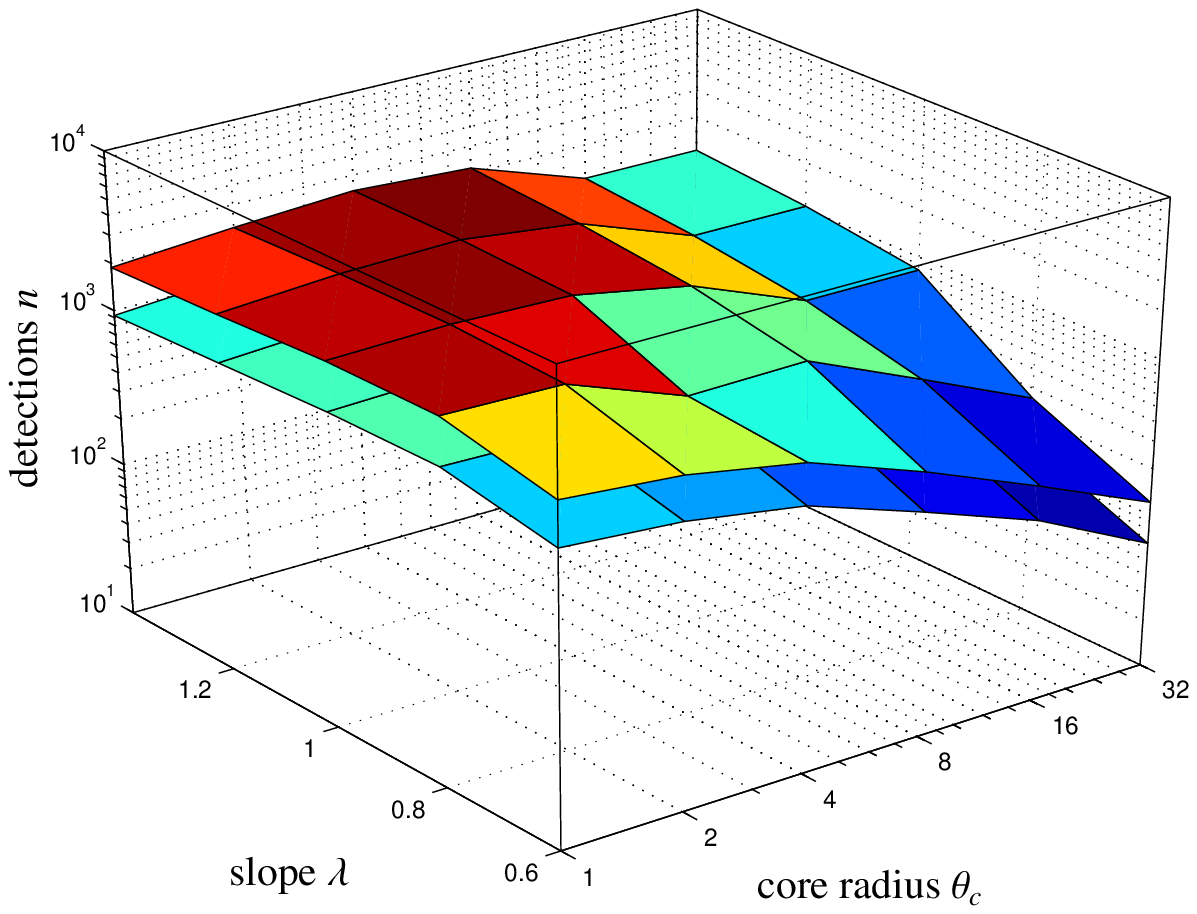}}\\
\resizebox{0.47\linewidth}{!}{\includegraphics{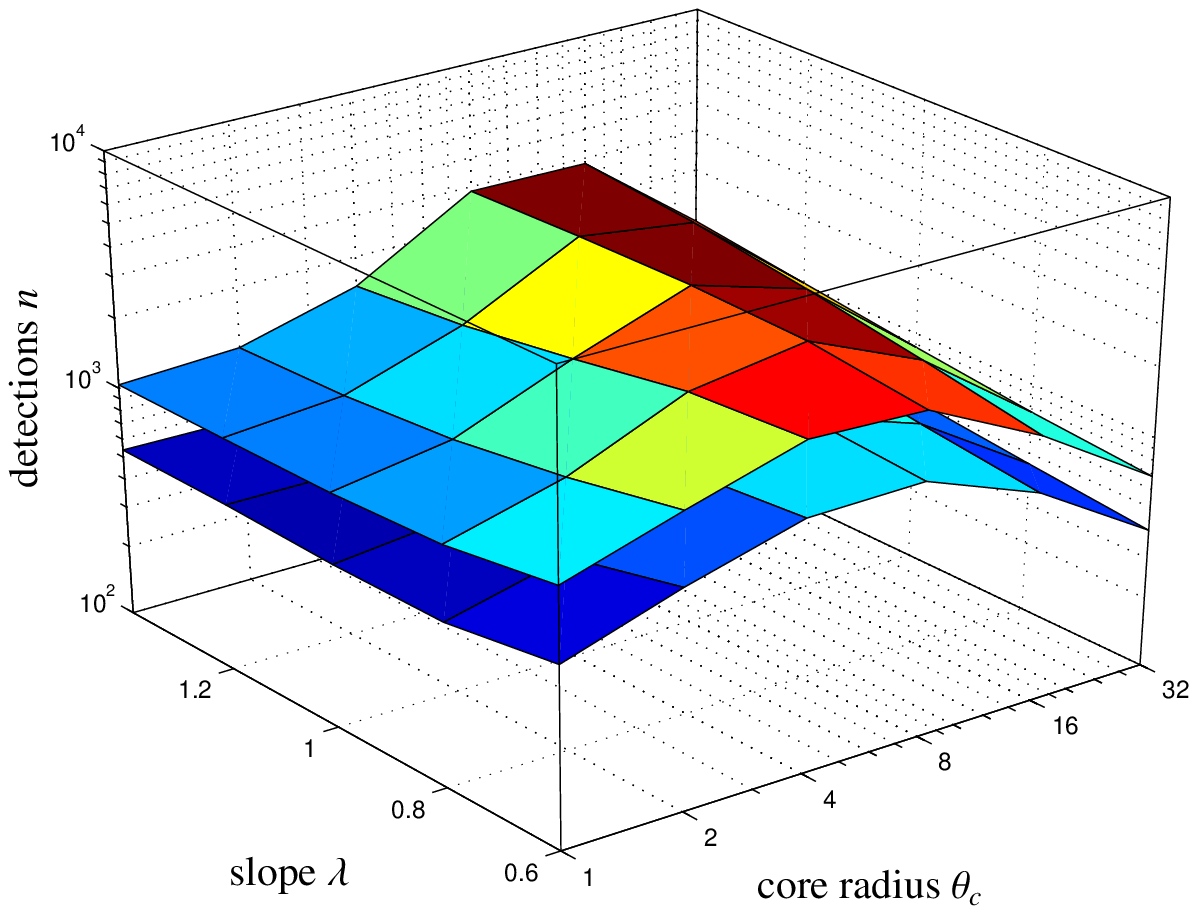}} & 
\resizebox{0.47\linewidth}{!}{\includegraphics{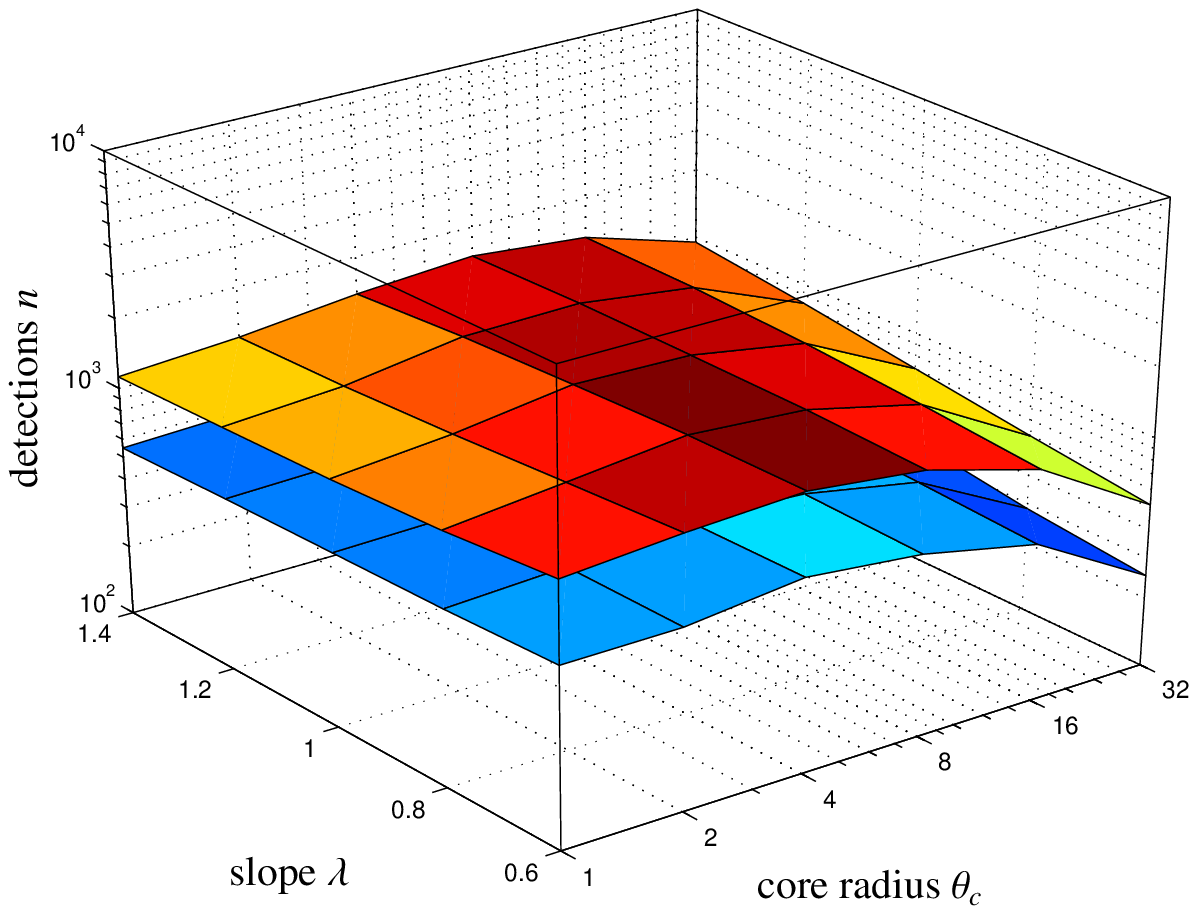}}
\end{tabular}
\end{center}
\caption{Number of detections $n(\theta_c,\lambda)$ as a function of both filter parameters core size $\theta_c$ 
and asymptotic slope $\lambda$, for the matched filter (top row) in comparison to the scale-adaptive filter (bottom row). 
The figure compares the number density following from a clean data set containing the CMB, the SZ-effects and instrumental noise 
(left column) with a data set containing all Galactic components in addition (right column). $n(\theta_c,\lambda)$ is given for the minimal signal strength $\mathcal{Y}_\mathrm{min}=3\times10^{-4}~\mathrm{arcmin}^2$ (upper plane) compared to $\mathcal{Y}_\mathrm{min}=10^{-3}~\mathrm{arcmin}^2$ (lower plane).}
\label{szobs_num_filter_2d}
\end{figure*}

The number density of clusters as a function of the King-profile's asymptotic slope $\lambda$ which the filters are optimised for is given in Fig.~\ref{szobs_num_filter_lambda}. The number of detections following from scale-adaptive filtering is relatively insensitive to particular choices of $\lambda$, whereas the matched filter yields a higher number of detections in the case of compact objects, irrespective of the noise components included in the analysis. Fig.~\ref{szobs_num_filter_2d} illustrates how the number of detections changes as a function of both $\theta_c$ and $\lambda$. It should be emphasised that none of the graphs depicted in Figs.~\ref{szobs_num_filter_theta}, ~\ref{szobs_num_filter_lambda} and \ref{szobs_num_filter_2d} shown the total number of detections for a given choice of $\theta_c$ and $\lambda$ and that graphs are not corrected for multiple detections of objects at more than $(\theta_c,\lambda)$-pair.

\subsection{Cluster population in the $M$-$z$-plane}\label{result_mzplane}
\begin{figure}
\resizebox{\hsize}{!}{\includegraphics{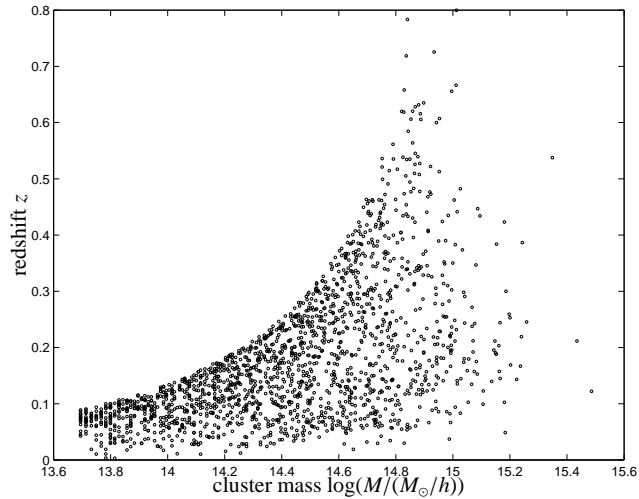}}
\caption{Population of clusters in the $\log(M)$-$z$-plane detected with the matched multifilter for the data set containing 
the CMB, instrumental noise and all Galactic foregrounds. The minimal signal strength was required to be $\mathcal{Y}_\mathrm{min}=10^{-3}~\mathrm{arcmin}^2$.}
\label{szobs_mzplane_matched}
\end{figure}

\begin{figure}
\resizebox{\hsize}{!}{\includegraphics{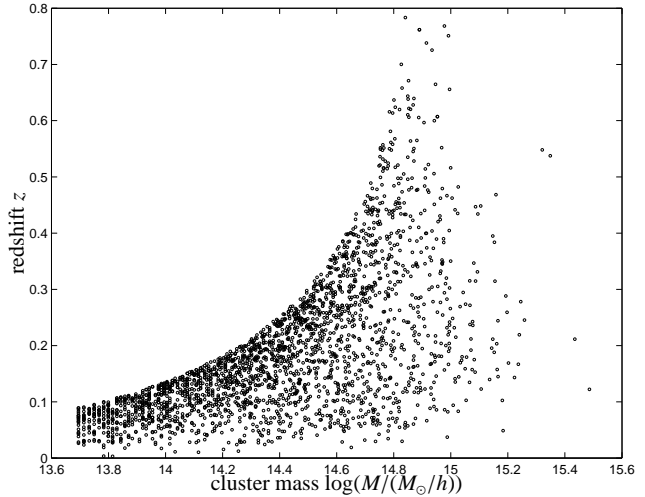}}
\caption{Population of clusters in the $\log(M)$-$z$-plane detected with the scale-adaptive multifilter. Here, the detections 
are given for a data set containing the CMB, instrumental noise and all Galactic foregrounds. All peaks exceed a minimial Comptonisation of $\mathcal{Y}_\mathrm{min}=10^{-3}~\mathrm{arcmin}^2$.}
\label{szobs_mzplane_scaleadaptive}
\end{figure}

Scatter plots describing the population of detectable clusters in the mass-redshift-plane are shown in 
Fig~\ref{szobs_mzplane_matched} for the matched filter and in Fig.~\ref{szobs_mzplane_scaleadaptive} for the scale-adaptive 
filter. The clusters populate the $\log(M)$-$z$-plane in a fairly well defined region. There are only few detections beyond redshifts of $z=0.8$, but the shape of the detection criterion suggests the existence of a region of low-mass low-redshift clusters which should be detectable but which are not included in the map construction. It is difficult to predict the SZ properties of low-mass clusters because many complications in the sector of baryonic physics come into play such as preheating, deviation from scaling laws and incomplete ionisation, which makes it difficult to predict the number of clusters missing in our analysis. Together with K. Dolag we prepared an auxiliary SZ-map from a gas-dynamical constrained simulation of the local universe that would fill in the gap and provide clusters with masses $M<5\times10^{13}M_\odot/h$ below redshifts of $z<0.1$ \citep{2005astro.ph..5258D}.

\begin{figure}
\resizebox{\hsize}{!}{\includegraphics{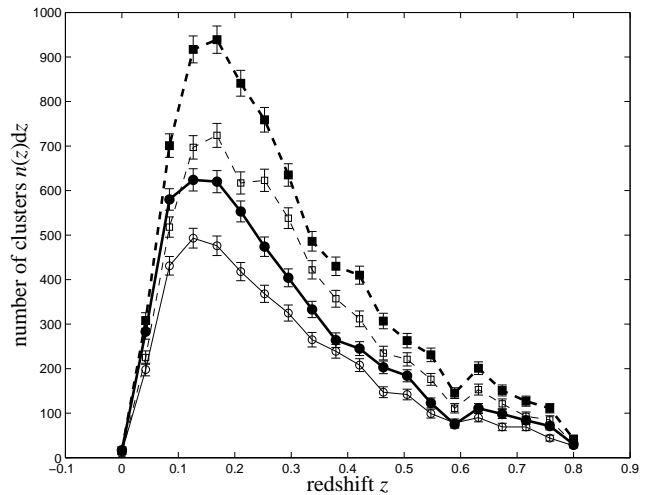}}
\caption{Distribution $n(z)\dd z$ of the detected clusters in redshift $z$, for the matched filter (solid 
line, circles) in comparison to the scale-adaptive filter (dashed line, squares). The figure compares detections in a clean 
data set containing the CMB, both SZ-effects and instrumental noise (thick lines, closed symbols) to a data set with all Galactic 
components in addition (thin lines, open symbols). Again $\mathcal{Y}_\mathrm{min}$ was set to $3\times10^{-4}~\mathrm{arcmin}^2$.}
\label{szobs_zdistro}
\end{figure}

Fig.~\ref{szobs_zdistro} gives the marginalised distribution in redshift $z$ of the cluster sample. The shape of the redshift distribution is determined by the competition of two effects: With increasing redshift $z$ the observed volume increases, but contrariwise, the number of massive clusters decreases as described by the Press-Schechter function and the SZ-signal becomes smaller proportional to $d_A^{-2}(z)$. Most of the clusters are observed at redshifts of $z\simeq0.2$ and the detection limit is reached at redshifts of $z\simeq0.8$. This applies to both filter algorithms and data sets alike.

\begin{figure}
\resizebox{\hsize}{!}{\includegraphics{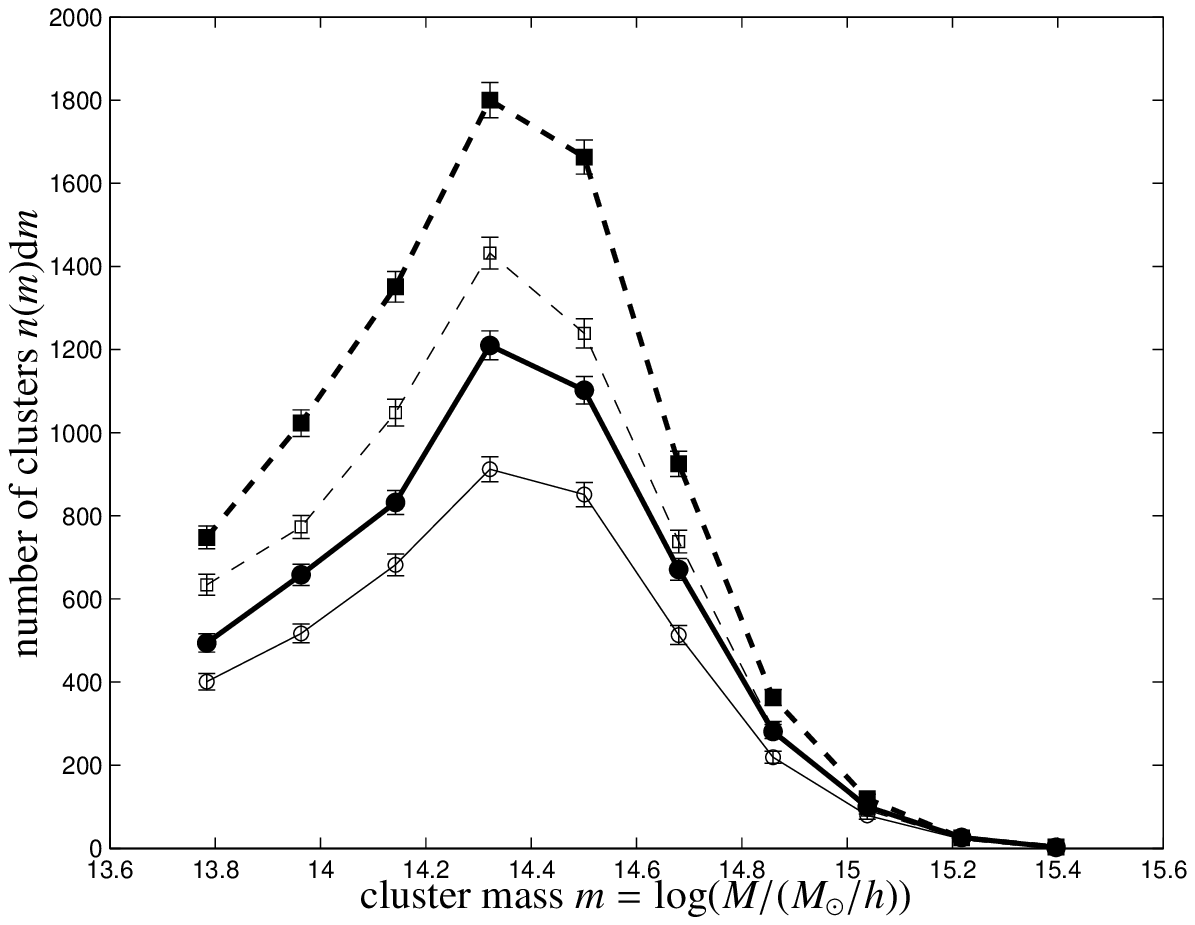}}
\caption{Distribution $n(m)\dd m$ of the detected clusters in logarithmic mass $m = \log(M/(M_{\solar}/h))$, for the matched 
filter (solid line, circles) in comparison to the scale-adaptive filter (dashed line, squares). Here, the distributions are 
given for a data set including only the CMB, both SZ-effects and instrumental noise (thick lines, closed symbols) in comparison 
to a data set containing moreover all Galactic foreground emission components (thin lines, open symbols). The minimal Comptonisation for spectroscopic confirmation was $\mathcal{Y}_\mathrm{min}=3\times10^{-4}~\mathrm{arcmin}^2$.}
\label{szobs_mdistro}
\end{figure}

\begin{figure}
\resizebox{\hsize}{!}{\includegraphics{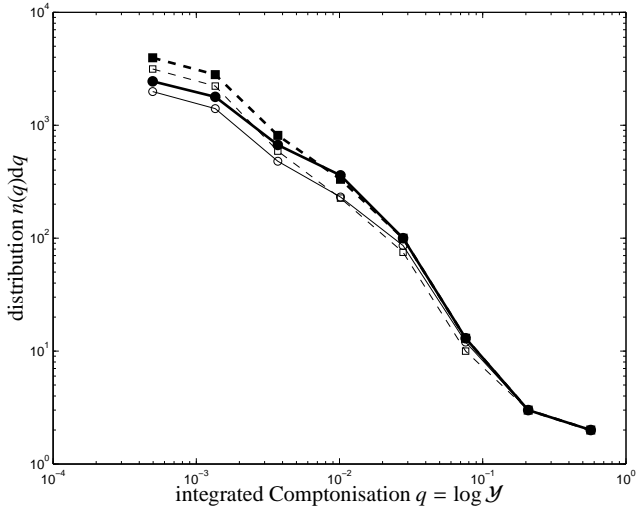}}
\caption{Distribution $n(q)\dd q$ of the logarithmic integrated Comptonisation, $q = \log(\mathcal{Y})$, for the matched 
filter (solid line, circles) in comparison to the scale-adaptive filter (dashed line, squares). Here, the distributions are 
given for a data set including only the CMB, both SZ-effects and instrumental noise (thick lines, closed symbols) in comparison 
to a data set containing moreover all Galactic foreground emission components (thin lines, open symbols).}
\label{szobs_inty}
\end{figure}

Fig.~\ref{szobs_mdistro} gives the marginalised distribution of the cluster's logarithmic mass $m = \log(M/(M_{\solar}/h))$. At high masses, both filtering schemes detect cluster reliably, but with decreasing mass, the filter algorithms start to show differences in their efficiency. The mass functions peak at a value of $2.5\times10^{14}M_\odot/h$, and decrease towards smaller values for the mass due to the decrease in number density of objects and smaller SZ-signal strengh $\mathcal{Y}$. Fig.~\ref{szobs_inty} gives the distribution of the cluster's Compton-$\mathcal{Y}$ parameter. The distribution is close to a power law as expected from virial estimates \citep[c.f. ][]{2004_szmap}, but at low Comptonisations, all distributions evolve shallower, which is due to the fact that clusters fail to generate a peak in the likelihood map exceeding the threshold value.

\subsection{Position accuracies}\label{result_sigma}
\begin{figure}
\resizebox{\hsize}{!}{\includegraphics{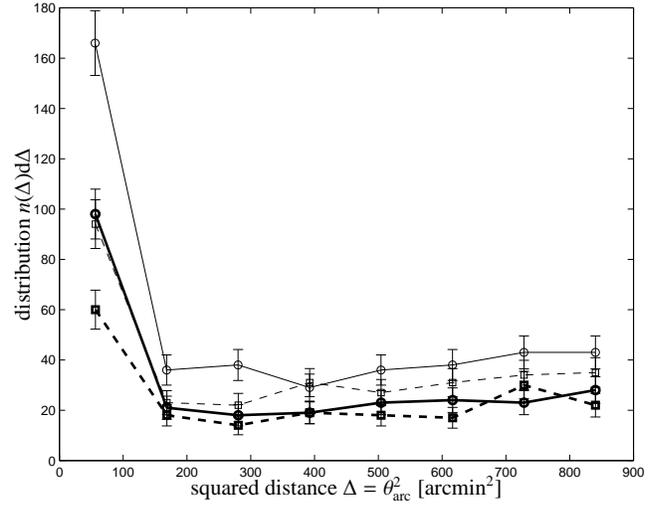}}
\caption{Distribution of the squared angular distance $\Delta=\theta_\mathrm{arc}^2$ between actual and reconstructed source position on a great circle, for the matched filter (solid line, circles) in comparison to the scale-adaptive filter (dashed line, squares). The figure compares detections above $4\sigma$ (thin lines) with detections above $5\sigma$ for clusters detected with the parameter $\theta_c=8\farcm0$. The clusters were required to generate a Comptonisation $\mathcal{Y}_\mathrm{min}$ exceeding $3\times10^{-4}~\mathrm{arcmin}^2$.}
\label{szobs_accuracies}
\end{figure}

A histogram of the deviations between actual and reconstructed cluster position is given by Fig.~\ref{szobs_accuracies}. The position accuracy is given in terms of the squared angular distance $\Delta=\theta_\mathrm{arc}^2$ because a uniform distribution would yield a flat histogram. The distribution is sharply peaked towards $\Delta=0~\mathrm{arcmin}^2$. A fraction of 50\% of all clusters are detected within $10^\prime$ from the nominal source position, but there is a tail in the distribution towards larger angular separations. For most of the clusters, this position accuracy is good enough for direct follow-up studies at X-ray wavelengths, but not good enough for optical observations.

As noticed in \citet{2002MNRAS.336.1057H}, it is not trivial to assign a peak in the filtered co-added map to an actual underlying cluster. Apart from this complication, the simulation itself introduces many sources of positional uncertainties: The distance shown in Fig.~\ref{szobs_accuracies} is the angular separation between the most-bound particle in the Hubble-volume simulation and the peak in the filtered map, i.e. apart from the misalignment of the peak in the filtered map relative to the maximum of the SZ-emission, there are misalignments of the SZ-template map relative to the cluster's barycenter and of the barycenter relative to the position of the most bound particle.

\subsection{Spatial distribution of \plancks SZ-cluster sample}\label{result_spatial_distrib}
\begin{figure}
\resizebox{\hsize}{!}{\includegraphics{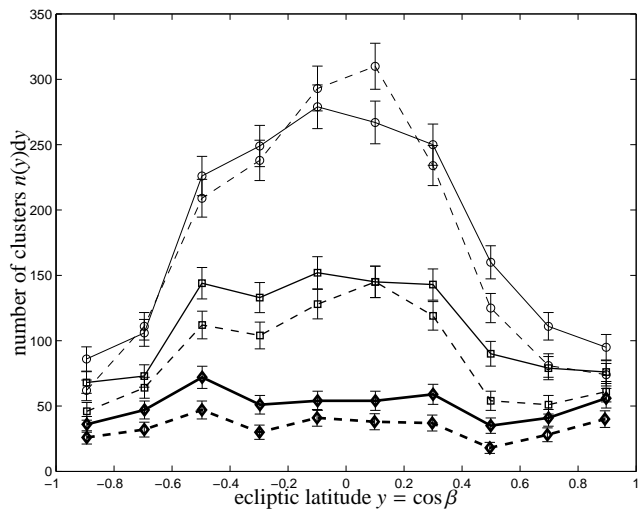}}
\caption{Number $n(y)\dd y$ of clusters as a function of ecliptic latitude $y=\cos\beta$, for the matched filter 
(solid line) in comparison to the scale-adaptive filter (dashed line). The figure compares the number of detected clusters as a function of ecliptic latitude for detection significances $>4.2\sigma$ (circles, thin lines), $>4.8\sigma$ (squares, medium lines) and $>6.0\sigma$ (diamonds, thick lines).}
\label{szobs_num_ecliptic}
\end{figure}

Fig.~\ref{szobs_num_ecliptic} shows the number density of clusters as a function of ecliptic latitude $y\equiv\cos\beta$. The figure states that the \planck~cluster sample extracted with the specific filters is highly non-uniform for low significance thresholds, where most of the clusters are detected on a belt around the celestial sphere, but gets increasingly more uniform with higher threshold values for the significance. This is due to the incomplete removal of low-$\ell$ modes in the filtered maps, which bears interesting analogies to the {\em peak-background split} \citep{1987Natur.330..451W, 1989MNRAS.237.1127C} in biasing schemes for linking galaxy number densities to dark matter densities: Essentially, the likelihood maps are composed of a large number of small-scale fluctuations superimposed on a background exhibiting a large-scale modulation. In regions of increased amplitudes due to the long-wavelength mode one observes an enhanced abundance of peaks above a certain threshold and hence an enhanced abundance of detected objects. 

\begin{figure}
\centering{\resizebox{\hsize}{!}{\includegraphics{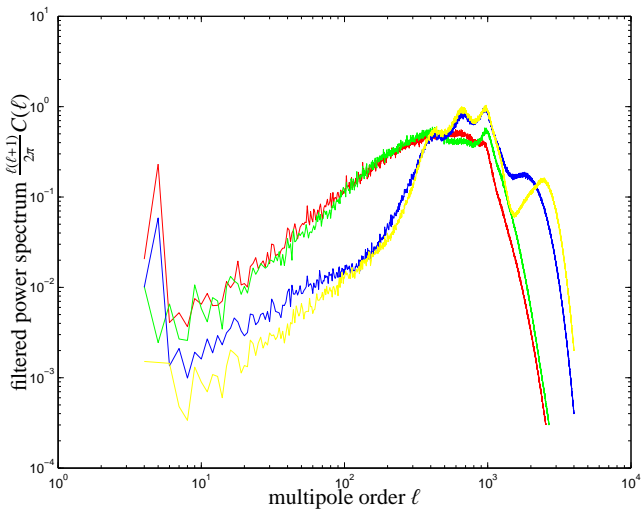}}}
\caption{Power spectra $C(\ell)$ of the filtered and co-added maps, where the filter kernels are derived for the parameters $(\theta_c,\lambda) = (4\farcm0,1.0)$, for the matched filter and data set containing the CMB realisation and instrumental noise (red line), for the matched filter and the data set containing all foregrounds in addition (green line), for the scale-adaptive filter and the data set including CMB fluctuations and instrumental noise (blue line) and for the scale-adaptive filter and the data set that contains all foregrounds in addition (yellow line).}
\label{szobs_powerspec_likelihood}
\end{figure}

As Fig.~\ref{szobs_powerspec_likelihood} indicates, the filtered and co-added maps do have large amplitudes for the hexadecupole which are certainly not in agreement with the near-Poissonian slope of $C(\ell)\propto\ell^2$ typical for a random distribution of small sources. The incomplete removal of low-$\ell$ modes shows that the assumptions about isotropy is violated on large scales and $C(\ell)$ ceases to be a fair description of the variance contained in the $a_{\ell m}$-coefficients. Clearly, this is a serious limitation to the spherical harmonic approach. In general, the low-$\ell$ fluctuations are more pronounced for extended objects, i.e. large $\theta_c$ and small $\lambda$, and they are stronger in the case of the matched filter compared to the scale-adaptive filter.

Similarly, detection significances near the detection threshold are inaccurate due to the long-wavelength modes. A way to remedy this would be to introduce local estimates of the mean and variance, for example by considering the average and the standard deviation of the amplitudes in an aperture with a few degrees in radius. One must keep in mind that in the filtered map, the signal is strong and likely to affect these two values. It should be emphasised, however, that the strongest signals exceeding values of $\simeq6\sigma$ are uniformly distributed over the celestial sphere.

\subsection{Detectability of the kinetic SZ-effect}\label{result_kinsz}
\begin{figure}
\resizebox{\hsize}{!}{\includegraphics{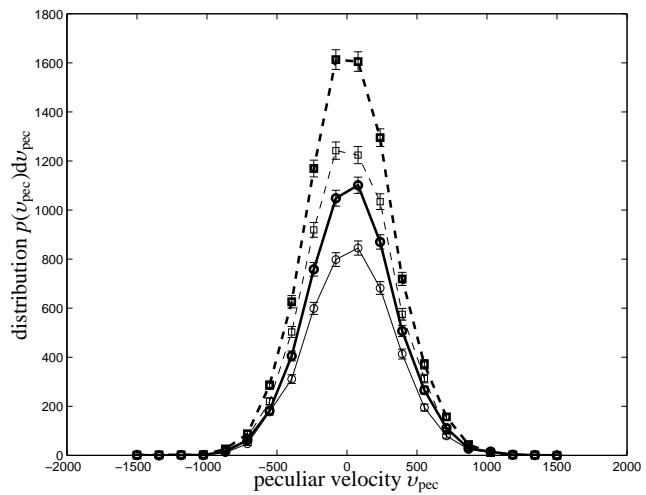}}
\caption{Number $n(\upsilon_\mathrm{pec})\dd \upsilon_\mathrm{pec}$ of clusters, for the matched filter (solid line, circles) in comparison to the scale-adaptive filter (dashed line, squares). Again, the detections in a data set containing the CMB, both SZ-effects and instrumental noise (thick lines, closed symbols) are compared to a data set containing all Galactic foregrounds in addition (thin lines, open symbols).}
\label{szobs_distro_pecvel}
\end{figure}

In this section, we give the distribution of peculiar velocities in \plancks SZ-cluster sample, which is an important guide for kinetic SZ-follow ups. As Fig.~\ref{szobs_distro_pecvel} indicates, the distribution of peculiar velocities are well approximated by a Gaussian with zero mean and standard deviation $\sigma_\mathrm{vel}\simeq 300~\mathrm{km}/\mathrm{s}$. For a dedicated search for the kinetic SZ-effect in \plancks SZ-cluster sample, velocities are drawn from this distribution, hence cluster bulk motions up to $300~\mathrm{km}/\mathrm{s}$ can be expected in 68\% of all cases and velocities in excess of $1000~\mathrm{km}/\mathrm{s}$ only for 11 to 16 objects, depending on the filtering scheme.

\section{Summary and discussion}\label{sect_summary}
The properties of the likelihood maps and of the cluster catalogues following from applying matched and scale-adaptive filtering to the simulated flux maps are characterised in detail. According to our simulation, \planck~can detect a number of $\simeq6000$ clusters of galaxies in a realistic observation with Galactic foregrounds (compared to over 8000 clusters if only the CMB and instrumental noise were present), which does not confirm the high numbers claimed by analytic estimates.

\begin{itemize}
\item{The noise properties of the filtered and co-added maps were examined in detail. It was found that the noise is very close to Gaussian after filtering, despite the fact that the initial flux maps had considerable anisotropic non-Gaussian features and despite the fact that the noise is highly structured and anisotropic on the cluster scale. Quantitatively, the variance of the filtered maps is smaller compared to the prediction based on the cross- and autocorrelation functions of the maps convolved with the filter. This discrepancy, which amounts to $\simeq 10\%$ is due to numerics, but has the effect that significances of peaks are slightly underestimated.The cluster detectability as a function of filter parameters showed that the matched filter performs better on compact objects, where its delivered significance depends strongly on the choice of $\lambda$. The scale-adaptive filter works well on extended objects and is relatively insensitive to $\lambda$.}

\item{The physical properties of the detected SZ-cluster sample made in terms of mass $M$, redshift $z$ and integrated Comptonisation $\mathcal{Y}$: The cluster population in the mass-redshift plane is fairly well defined, and the marginalisation over the mass resulted in most of the clusters being detected at redshifts of $z\simeq0.2$, where the distribution starts decreasing to values of $z\simeq0.8$, where no clusters are detected. The distribution of detected SZ-clusters in mass $M$ confirmed that the high-mass end of the Press-Schechter function is well sampled, that most of the clusters detected have masses $\simeq2.5\times10^{14}M_\odot/h$ and that clusters of lower mass are increasingly difficult to detect.}

\item{The position accuracy is better than $10^\prime$ in half of the cases, which is sufficient for X-ray follow-up studies, but the distribution exhibits a tail towards high discrepancies between the cluster position and the position of the peak in the likelihood map.}

\item{The investigation of the spatial distribution, especially in ecliptic latitude showed that the distribution of clusters gets increasingly uniform with increasing detection threshold. This is due to the fact that the filtered and co-added maps exhibit long-wavelength variations due to insufficient filtering at low multipoles.}

\end{itemize}

The simulation as presented here has a number of shortcomings that may affect SZ-predictions:
\begin{itemize}
\item{It was assumed for reasons of computational feasibiliy that all Galactic foregrounds had isotropic spectral properties. While this is an excellent approximation for the CMB, Galactic components can be expected to exhibit spatially varying spectral properties. For example, the spectral index of the Galactic synchrotron emission is likely to change with the propeties of the population of relativisitic electrons and the magnetic field and the spectrum of thermal dust changes with the dust temperature. The filter construction as it is would be applicable to those cases as well despite the fact that at fixed angular scale $\pi/\ell$, the cross power spectrum $C_{\nu_i\nu_j}(\ell)$ between frequencies $\nu_i$ and $\nu_j$ ceases to be a good description of the cross-variance contained in the $a_{\ell m}(\nu_i)$-coefficients.}

\item{Another related point worth mentioning is the approximate derivation of the covariance matrix from the $a_{\ell m}(\nu_i)$-coefficients with the relation $C_{\nu_i\nu_j}(\ell)=(2\ell+1)^{-1}\sum_{m=-\ell}^{+\ell} a_{\ell m}(\nu_i)a_{\ell m}^*(\nu_j)$ for the computation of matched and scale-adaptive filter kernels. This formula yields an unbiased estimate of the power spectrum $C_{\nu_i\nu_j}(\ell)$ only in the case of Gaussian random fields, which is certainly not the case if the Galaxy is included in the analysis. It would be more appropriate to derive the value of $C_{\nu_i\nu_j}(\ell)$ that maximises the likelihood of describing the data, as described in \citet{1998PhRvD..57.2117B}, and to use this power spectrum for the derivation of filter kernels.}

\item{We did not include ICM physics beyond adiabaticity. Cooling processes in the centres of clusters give rise to cool cores, which can be shown to boost the line-of-sight Comptonisation $y$ by a factor of $\sim2-3$. The volume fraction occupied by such a cool core is very small compared to the entire cluster and hence the total integrated Comptonisation $\mathcal{Y}$ does not change significantly. For a low-resolution observatory like \planck, the primary observable is $\mathcal{Y}$, and for that reason, SZ-observations carried out with \planck~should not be affected by cool cores. A further complication is the existence of non-thermal particle populations in the ICM, but their contribution to the SZ-flux modulation is very small.}

\item{There is a serious issue concerning completeness. The population of detections in the $M$-$z$ plane suggests that low-mass clusters at redshifts $z<0.1$ should be detectable for \planck. This particular region of the $M$-$z$-plane is not covered by the SZ-map construction, but \planck~would certainly add detections in this particular region of the $M$-$z$-plane. The SZ-maps of the local universe provided by \citet{2005MNRAS.361..753H} fill in this gap of low-mass, low-redshift systems and can be combined with the SZ-maps covering the Hubble volume described in this work.}

\item{Extragalactic point sources were excluded from the analysis due to poorly known spectra and clustering properties. In the simplest case of homogeneously distributed sources, there is a Poisson fluctuation in the number of point sources inside the beam area, which causes an additional noise component with power spectrum $C(\ell)\propto\ell^2$ similar to uncorrelated pixel noise. If these sources have similar spectral properties, they could be efficiently suppressed by the linear combination of observations at different frequencies.}

\item{We did not attempt to simulate effects arising in the map making process and complications due to the $1/f$-noise. So far it has not been investigated how well small structures can be reconstructed from time-ordered data streams. The map-making algorithms are chiefly optimised to yield good reconstructions of the CMB fluctuations by recursively minimising the noise, but to our knowledge the reconstruction of compact objects like SZ-clusters or minor planets has not been simulated for these algorithms. At the cluster scale, the dominating noise component is uncorrelated pixel noise, so that the contamination by $1/f$-noise does not play a role on these scales.}

\item{Gaps in the data are a serious issue for the filtering schemes: Blank patches in the observed sky cause the power spectra $C_{\nu_1\nu_2}(\ell)$ at different multipole order $\ell$ to be coupled due to convolution with the sky window function. This is due to the fact that the $Y_{\ell m}(\theta,\phi)$-basis ceases to be an orthonormal system if the integration can not be carried out over the entire surface of the celestial sphere. Because the linear combination coefficients are determined separately for each multipole moment $\ell$ from the inverse of the covariance matrix $C_{\nu_1\nu_2}(\ell)$, correlations between the covariance matrices at differing $\ell$ are likely to yield an insufficient reduction of foregrounds.}

\item{Galactic templates, especially the carbon monoxide map and the free-free map, are restricted to relatively low values in $\ell$ and do not extend to high multipoles covered by \planck. For that reason, foreground subtraction at high values of $\ell$ is likely to be more complicated in real data. Furthermore, one should keep in mind that the frequencies above 100~GHz are a yet uncharted territory and although the existence of an unknown Galactic emission component seems unlikely, the extrapolation of fluxes by two to three orders of magnitude in frequency may fail.}
\end{itemize}

The capability of \planck~to detect SZ-clusters has been the subject of many recent works, pursuing 
analytical \citep{1997A&A...325....9A, 2002hzcm.conf...81D, 2001A&A...370..754B, 2002MNRAS.335..984M} as well as 
semi-analytical \citep{2000ApL&C..37..341S, 2002MNRAS.336.1057H, 2001MNRAS.325..835K, 2002MNRAS.336.1351D, 2003MNRAS.338..765H, 2004astro.ph..6190G} and numerical approaches \citep{2003ApJ...597..650W}. 
\begin{itemize}
\item{\citet{1997A&A...325....9A} use analytical $\beta$-profiles, an $M$-$T$ relation from $n$-body data and the Press-Schechter function for generating square SZ-maps with side length $\sim12\degr$. To this map they superimposed CMB fluctuations, Galactic foregrounds and instrumental noise. From this data they recovered the SZ-signal by multifrequency Wiener-filtering outlined by \citet{1999MNRAS.302..663B}. They predict a total number of 7000 clusters with integrated Comptonisations $\mathcal{Y}\gsim9\times10^{-4}~\mathrm{arcmin}^2$ and $10^4$ objects at $\mathcal{Y}\gsim5\times10^{-4}~\mathrm{arcmin}^2$. These numbers slightly exceed our results.}

\item{The paper by \citet{2001A&A...370..754B} and the related work by \citet{2002MNRAS.335..984M} take a purely analytical approach with spherically symmetric $\beta$-profiles for describing the SZ-morphology and rely on the Press-Schechter function and an $M$-$T$ relation from numerical data for predicting the SZ-signal of clusters. They incorporate the effect of the finite instrumental resolution and require the integrated Comptonisation $\mathcal{Y}$ to exceed the value of $3\times10^{-4}~\mathrm{arcmin}^2$. The total number of detectable clusters is stated to be $10^4$, which again slightly exceeds our findings, but the distribution of cluster masses $M$ and the distribution of detectable clusters in redshift $z$ is very similar to the results presented in this paper. The redshift distribution peaks at a very similar value, but extends to larger redshifts beyond $z\simeq0.8$.}

\item{The papers written by \citet{2001ApJ...552..484S} and \citet{2002MNRAS.336.1057H}, who developed the concept of matched and scale-adaptive multifiltering based on an extremal principle for flat topologies and Fourier-decomposition as the harmonic system, concentrate mainly on filter construction. They employ analytic SZ-profiles and describe the instrumental noise as uncorrelated Gaussian pixel noise, but consider the entire spectrum of Galactic foregrounds. \citep{2002MNRAS.336.1057H} advocate a number of $\simeq10^4$ clusters to be detectable by \planck, which they estimate by extrapolating the average number of detections in simulated $12\degr$-wide patches to the entire celestial sphere, while restricting themselves to higher Galactic latitudes of $\left|b\right|\gsim19\degr$ ($f_\mathrm{sky}=2/3$). Similarly, \citet{2002MNRAS.336.1351D} applies a Bayesian non-parametric method to the same data and finds a total number of $9\times10^3$ clusters at Galactic latitudes of $\left|b\right|\gsim12\degr$ ($f_\mathrm{sky}=0.8$). Thus, both analyses are quite comparable with our approach concerning their estimated number of detections, while yielding cluster catalogues that contain slightly more entries.}

\item{In the study by \citet{2001MNRAS.325..835K} the SZ-population was modelled using the Hubble-volume simulation as a cluster catalogue. The main aim is the difference of SZ-catalogues delivered by \planck~in the $\Lambda$CDM cosmology compared to the $\tau$CDM model. The expected SZ-signal was derived based on an $M$-$T$-relation and they include an instrumental description including finite resolution and frequency response. By requiring a cluster to generate an integrated Comptonisation exceeding the value of $3\times10^{-4}~\mathrm{arcmin}^2$ in an area defined by the virial radius, they find a total number of $5\times10^4$ clusters of which a fraction of 1 percent is spatially resolved. The limiting redshift is stated to be $z\simeq1.5$, while the distribution in redshift peaks at a comparatively large value of $z\simeq0.3\ldots0.4$. In comparison, their cluster catalogue exceeds ours by a factor of a few, which is likely due to the fact that they neither consider instrumental noise nor foregrounds, but concentrate rather on the fluctuating SZ-background alone as the main source of noise.}

\item{The Bayesian approach by \citet{2003MNRAS.338..765H} focuses mainly on the problem of peak finding and shows that the method they investigate can be readily applied to \planck-data. They consider a very simplified SZ-observation with \planck-characteristics at a single frequency, use analytic profiles, neglect all Galactic foregrounds and include only the fluctuating CMB and instrumental noise as noise sources. Consequently, they do not give astrophysical properties of the SZ-clusters their method is able to find, but it should be emphasised that the quantification of a peak height in terms of a Bayesian likelihood is far preferrable to our quantification in terms of a statistical significance.}

\item{In contrast, the filter scheme employed in the paper by \citet{2004astro.ph..6190G} is the powerful harmonic-space maximum entropy method introduced by \citet{2002MNRAS.336...97S}, which is primarily optimised for component separation rather than the detection of individual objects. The SZ-signal they put into the simulation is determined from scaling relations and uses spherically symmetric analytic profiles. Including an accurate description of \plancks instrumentation and Galactic foregrounds, they find a total number of up to $1.1\times10^4\ldots1.6\times10^4$ clusters depending the $M$-$T$-relation for the choice $\sigma_8=0.9$, while these numbers decrease by $\simeq35$\% for a $20\degr$-wide Galactic cut. Their distribution in redshift $z$ is quite similiar in shape compared to ours - neither of us finds high-redshift clusters beyond $z=1$, but their distribution falls off slower with increasing redshift $z$. A grand result is their extraction of the SZ power spectrum, which our analysis due to its focus on the detection of individual peaks is not able to deliver. It should be kept in mind, however, that the component separation method, despite its prowess, assumes prior approximate knowledge of the emission component's power spectra, which are only partially available at HFI-frequencies above $\nu=100$~GHz.}

\item{\citet{2003ApJ...597..650W} puts emphasis on using a hydrodynamical simulation of structure formation, albeit on a small scale, but including non-collapsed objects. He includes neither CMB fluctuations nor Galactic foregrounds, but chooses a high threshold value for a simple flux criterion for detection. The signal amplification strategy is smoothing with a Gaussian kernel and linear combination of maps. From the number of detections on a small patch of the sky he coarsely estimates the extrapolated number of $10^3\ldots10^4$ clusters without giving details on their astrophysical characteristics, but concentrates rather on the observability in follow-up studies. Extending this investigation, \citet{2003ApJ...586..723S} have considered a range of future CMB experiments including \planck~and modelled SZ-observations on the basis of a large $n$-body simulation with a semianalytic receipe for the gas distribution inside dark matter structures. They restrict themselves to high Galactic latitudes and include CMB fluctuations as well as (Gaussian) instrumental noise. Signal extraction is performed by applying a matched filter algorithm. They quantify catalogue completeness and emphasise complications in peak finding.}
\end{itemize}

In conclusion, the simulation presented in this paper demonstrates the abilities of \planck~with respect to detecting Sunyaev-Zel'dovich clusters of galaxies even in the presence of anisotropic non-Gaussian noise components with complicated spectral dependences. Despite the fact that the high number of detections claimed by analytical estimates needs to be adjusted, it was shown that our results support the expectations on \plancks cluster sample and that the numerical tools for analysing the cross- and autocorrelation properties of all \planck~channels and for filtering the data work reliably up to the high multipoles of $\ell=4096$ considered here. The \planck~catalogue of SZ-clusters of galaxies will surpass X-ray catalogues \citep[e.g. the REFLEX catalogue compiled by][ on the basis the {\em Rosat} all-sky survey]{2004A&A...425..367B} in numbers as it reaches deeper in redshift and is able to detect low-mass systems. It will contribute to the determination of cosmological parameters related to structure formation and dark energy, and shed light on baryonic physics inside clusters of galaxies.

\section*{Acknowledgements}
The help of Martin Reinecke in enhancing the \planck-simulation tools, in adding custom changes and his support in interfacing our software with the \planck-simulation tools is greatly appreciated. We would like to thank Christoph Pfrommer for discussions.

\appendix

\bibliography{bibtex/aamnem,bibtex/references}
\bibliographystyle{mn2e}

\appendix

\bsp

\label{lastpage}

\end{document}